\documentclass[
 aps,
 pra,
 reprint,
 superscriptaddress,
 nofootinbib,
 floatfix
]{revtex4-2}

\usepackage[T1]{fontenc}
\usepackage{microtype}
\usepackage{amsmath,amssymb,amsthm,mathtools,bm}
\usepackage[colorlinks=true,allcolors=blue]{hyperref}

\newtheoremstyle{apsresult}
 {6pt}{6pt}{\normalfont}{}%
 {\itshape}{.}{0.5em}{}%
\theoremstyle{apsresult}
\newtheorem{theorem}{Theorem}

\newtheorem{lemma}{Lemma}
\newtheorem{corollary}{Corollary}
\newtheorem{definition}{Definition}

\newcommand{\Tr}{\operatorname{Tr}}
\newcommand{\rank}{\operatorname{rank}}
\newcommand{\ran}{\operatorname{ran}}
\newcommand{\supp}{\operatorname{supp}}
\newcommand{\Span}{\operatorname{span}}
\newcommand{\Fr}{\operatorname{Fr}}
\newcommand{\SO}{\operatorname{SO}}

\newcommand{\R}{\mathbb{R}}
\newcommand{\I}{\mathbb{I}}
\newcommand{\calE}{\mathcal{E}}
\newcommand{\calP}{\mathcal{P}}
\newcommand{\calT}{\mathcal{T}}
\newcommand{\calZ}{\mathcal{Z}}
\newcommand{\Bbar}{\overline{\mathbb B}^{3}}
\newcommand{\bn}{\bm n}
\newcommand{\br}{\bm r}
\newcommand{\bu}{\bm u}
\newcommand{\bv}{\bm v}
\newcommand{\bw}{\bm w}
\newcommand{\bmom}{\bm m}
\newcommand{\bsigma}{\bm\sigma}
\newcommand{\bone}{\bm 1}
\newcommand{\dd}{\mathrm d}
\newcommand{\ee}{\mathrm e}
\newcommand{\ii}{\mathrm i}
\newcommand{\TF}{\operatorname{TF}}

\begin{document}

\title{Global Minimax Readout of a Qubit Direction}

\author{Abbas Taherpour}
\email{taherpour@qu.edu.qa}
\affiliation{IIPL, 
Electrical Engineering Department, Qatar University,
P.O. Box 2713, Doha, Qatar.
}

\author{Amirhossein Taherpour}
\email{at3532@columbia.edu}
\affiliation{
Department of Electrical Engineering,
Columbia University,
New York, NY 10027, USA.
}

\author{Tamer Khattab}
\email{tkhattab@ieee.org}
\affiliation{IIPL, 
Electrical Engineering Department, Qatar University,
P.O. Box 2713, Doha, Qatar.
}

\begin{abstract}
We determine the exact worst-direction Fisher-information cost of using a
parameter-independent readout to estimate an unknown qubit direction at known
Bloch-vector length $\eta$.  Every fixed-local architecture, including
recorded classical randomization, heterogeneous single-copy measurements, and
arbitrary outcome spaces, reduces exactly to a zero-barycenter probability
measure on the Bloch ball.  For every full-rank qubit and every trace-balanced
spectral Fisher loss, the resulting minimax problem is rigid: the unique
optimal aggregate design is the spin-coherent Haar positive-operator-valued
measure (POVM).  For $N$ copies, inverse-Fisher $A$ loss has the exact value
$2/[Nf(\eta)]$, where $
 f(\eta)=
 \frac{2\eta-(1-\eta^2)\log[(1+\eta)/(1-\eta)]}{4\eta}$. This uniqueness has an immediate finite-readout consequence.  No
finite-support measurement attains the unrestricted mixed-state optimum, while
at the smallest globally regular support the tetrahedral symmetric
informationally complete (SIC) measurement is uniquely $A$- and $D$-minimax,
with exact worst-direction values.  Relaxing the fixed-readout constraint
separates the asymptotic resources: one-way local operations and classical
communication (LOCC), unrestricted LOCC, and separable measurements have
$A$-loss coefficient $4/\eta^2$, whereas collective measurements attain
$2(1+\eta)/\eta^2$.  We further classify the rigidity conditions for unequal
contrasts and show that Haar uniqueness survives at the nonregular pure-state
endpoint.
\end{abstract}

\maketitle

\section{Introduction}
\label{sec:introduction}

Quantum-estimation theory usually optimizes a measurement in a neighborhood
of a known operating point.  A sensor deployed before the parameter is known
faces a different task: one parameter-independent readout must remain
informative over the entire parameter space.  In that setting a local quantum
Cram\'er--Rao bound does not by itself identify the best fixed measurement,
nor does it quantify the statistical price of forbidding parameter-dependent
retuning
\cite{Mukhopadhyay2025,MukhopadhyayParisBayat2025,Rubio2024,
Boeyens2025}.

We study this distinction for the direction $\bn\in S^2$ of a qubit with
known Bloch-vector length $0<\eta\leq1$.  The measurement is selected without
knowledge of $\bn$, and performance is evaluated by the worst directional
Fisher loss.  Our baseline fixed-local architecture is deliberately broad:
the single-copy positive-operator-valued measures (POVMs) may differ from copy
to copy, settings may be randomized or scheduled, outcome spaces may be
discrete or continuous, and every setting label and outcome is retained.
Thus recorded classical randomization is part of the statistical experiment
rather than unobserved noise \cite{DAriano2005}.  We exclude only
outcome-dependent feedforward within the block and quantum measurements
coupling distinct copies.

Symmetry makes the spin-coherent Haar POVM an immediate candidate, but it does
not settle the problem addressed here.  Covariant-measurement and invariant
decision-theory arguments can produce a covariant optimizer without excluding
additional asymmetric optimizers \cite{Davies1978,Holevo2011}.  Likewise,
Fisher-symmetric and informationally complete measurements describe important
isotropy and reconstruction properties
\cite{Renes2004,Scott2006,Zhu2014,Li2016,ZhuHayashi2018}, but those properties
do not by themselves solve a fixed-radius, worst-direction minimax problem.

Related formulations illuminate complementary aspects of measurement design.
Suzuki formulated qubit state estimation in the language of optimal design of
experiments and derived $A$-, $D$-, and $E$-optimal designs for the local,
pointwise design problem \cite{Suzuki2021OptimalDesign}.  Minimax quantum-state
estimation has also been studied at the level of estimator risk under Bregman
divergences; in the qubit case, spherical $2$-designs arise as minimax
measurements in that setting \cite{Quadeer2019Minimax}.  The objective here is
different from both.  We optimize the spectrum of the ordinary Fisher
information of a single parameter-independent measurement experiment, at
fixed Bloch radius, against the worst unknown direction, and we classify all
measurements that attain the optimum.  Recent work on global sensing,
randomized multiparameter measurements, and the Fisher geometry of fixed
informationally complete measurements further motivates such
state-independent readout design
\cite{MukhopadhyayParisBayat2025,ZhouChen2026,Saini2025}.

The central result is a rigidity theorem.  We first show that every admissible
fixed-local architecture is represented, for Fisher-information purposes, by
a single zero-barycenter probability measure on the Bloch ball.  A rotational
trace identity then gives the sharp minimax value for every trace-balanced
spectral loss.  Attaining that value forces all contributing effects to be
rank one and the Fisher field to be isotropic at every direction.  The
remaining equality condition is a spherical convolution equation.  Its
higher-degree harmonic multipliers are nonzero, so mass and barycenter
constraints eliminate the remaining degrees and force the aggregate design
measure itself to be Haar. The sharp trace bound together with the covariant Haar construction establishes
Haar optimality, whereas the equality analysis proves something stronger: the
spin-coherent Haar design is the unique aggregate optimizer.

This rigidity has an immediate finite-resource consequence.  Because equality
requires the Haar measure itself, no finite-support readout can attain the
unrestricted mixed-state optimum.  We obtain a support-dependent lower bound
for every finite alphabet and solve exactly the smallest globally regular
case.  At four atoms, equality in the universal atom-weight bound forces equal
weights and regular-tetrahedral geometry; an exact calculation of the Fisher
field then shows that the tetrahedral symmetric informationally complete
(SIC) measurement is uniquely $A$- and $D$-minimax and identifies its four
antipodes as the only worst directions.  The finite-readout problem is
therefore not merely an approximation question: Haar rigidity creates a
strict resource separation between continuous and finite fixed readouts.

A second consequence is a clean comparison of measurement architectures.
For $0<\eta<1$, the asymptotic $A$-loss coefficients satisfy
\begin{equation}
 \frac{2}{f(\eta)}
 >
 \frac{4}{\eta^2}
 >
 \frac{2(1+\eta)}{\eta^2},
 \label{eq:intro-hierarchy}
\end{equation}
corresponding respectively to fixed-local readout, adaptive local operations
and classical communication (LOCC) or separable readout, and collective
measurement.  The first gap is the price of committing to one globally fixed
measurement before the direction is known.  Feedforward removes this
global-readout cost asymptotically while retaining the local multiparameter
incompatibility cost, and collective measurement removes the remaining
incompatibility.  For the collective upper bound we give an explicit
parameter-independent construction: the Schur spin sector is first resolved,
and a spin-coherent covariant POVM is then performed within each sector.  Its
ordinary Fisher field is isotropic and attains the collective coefficient
asymptotically.

The rigidity mechanism also survives two useful stress tests.  With unequal
single-copy contrasts, isotropy separates into two independent harmonic
cancellation conditions.  One contrast class must be Haar in aggregate, two
distinct classes must each be Haar separately, while three or more classes
permit exact anisotropic cancellation between groups.  At the pure-state
endpoint $\eta=1$, antipodal zero-probability outcomes make the ordinary
Fisher model nonregular and remove the higher harmonics from the scalar trace
kernel.  Haar uniqueness nevertheless survives: the traceless tensor
condition, expressed through a distributional trace-free Hessian, still
annihilates every higher harmonic.

The paper is organized as follows.  Section~\ref{sec:design} reduces fixed
readouts to zero-barycenter design measures on the Bloch ball.
Section~\ref{sec:minimax} establishes the pointwise benchmark and proves Haar
minimax rigidity.  Section~\ref{sec:finite} derives the finite-support
separation and the exact tetrahedral optimum.  Section~\ref{sec:robustness}
treats unequal contrasts and the pure-state endpoint.
Section~\ref{sec:architectures} compares fixed-local, LOCC, separable, and
collective measurement architectures.  Detector-degree classifications,
deterministic-calibration topology, and the complete proofs are collected in
the appendices.

\section{Fixed readout as a design problem}
\label{sec:design}

A homogeneous block contains $N$ labeled qubits in
\begin{align}
 \rho_{\bn}^{[N]}&=\rho_{\bn}^{\otimes N},
 \label{eq:model-block}\\
 \rho_{\bn}&=\frac12\left(\I+\eta\,\bn\cdot\bsigma\right),
 \qquad \bn\in S^2.
 \label{eq:model}
\end{align}
Sections~\ref{sec:design}--\ref{sec:robustness} assume $0<\eta<1$ unless the
pure endpoint is stated explicitly.  Let
\begin{equation}
 P_{\bn}=I_3-\bn\bn^T
 \label{eq:tangent-projector}
\end{equation}
be the orthogonal projector onto $T_{\bn}S^2$.  For a recorded experiment
$\calE$ and tangent vectors $\bv,\bw\in T_{\bn}S^2$, its Fisher operator is
\begin{equation}
 \langle\bv,F_{\calE}(\bn)\bw\rangle
 =
 \int
 \frac{
  \dd p_{\calE}(y|\bn)[\bv]\,
  \dd p_{\calE}(y|\bn)[\bw]
 }{
  p_{\calE}(y|\bn)
 }.
 \label{eq:fisher-def}
\end{equation}
The integral includes sums over discrete outcomes.  For $0<\eta<1$, every
nonzero effect has strictly positive probability, so the ordinary Fisher
model is regular \cite{BraunsteinCaves1994}.

A fixed-local architecture is specified by a parameter-independent recorded
variable $\lambda\in\Lambda$, with law $\xi$.  Conditional on $\lambda$, node
$j$ is measured by an arbitrary qubit POVM $M_{\lambda j}$; all local outcomes
and $\lambda$ are retained.  This includes fixed POVMs, finite or continuous
setting randomization, heterogeneous local POVMs, and deterministic schedules
through their empirical setting frequencies.  It excludes outcome-dependent
retuning and joint measurements across copies.

We optimize a class of losses for which a trace bound becomes sharp exactly
at an isotropic Fisher matrix.

\begin{definition}[Trace-balanced spectral loss]
\label{def:trace-balanced}
A continuous symmetric function
$\ell:(0,\infty)^2\to\R$ is trace balanced when, with
\begin{equation}
 \psi_{\ell}(t)=\ell(t/2,t/2),
 \label{eq:psi-def}
\end{equation}
the following conditions hold:
\begin{enumerate}
 \item $\psi_{\ell}$ is strictly decreasing;
 \item
 \begin{equation}
  \ell(x,y)\geq\psi_{\ell}(x+y),
  \label{eq:trace-balance}
 \end{equation}
 with equality if and only if $x=y$;
 \item $\ell(x_k,y_k)\to+\infty$ whenever
 $\min\{x_k,y_k\}\to0^+$ while $x_k+y_k$ remains bounded.
\end{enumerate}
For a positive tangent Fisher operator with eigenvalues
$\lambda_1,\lambda_2$, write
\begin{equation}
 L_{\ell}(F)=\ell(\lambda_1,\lambda_2),
 \label{eq:spectral-loss}
\end{equation}
and set $L_{\ell}(F)=+\infty$ when $F$ is singular.
\end{definition}

The standard examples are
\begin{align}
 \ell_A(x,y)&=x^{-1}+y^{-1},
 &\psi_A(t)&=\frac4t,
 \label{eq:A-loss}\\
 \ell_D(x,y)&=(xy)^{-1/2},
 &\psi_D(t)&=\frac2t,
 \label{eq:D-loss}\\
 \ell_E(x,y)&=\max\{x^{-1},y^{-1}\},
 &\psi_E(t)&=\frac2t.
 \label{eq:E-loss}
\end{align}
The $A$ loss is $\Tr(F^{-1})$.  These criteria optimize Fisher experiments;
they do not by themselves assert a finite-sample global risk theorem.

Let $\Bbar=\{\br\in\R^3:\|\br\|\leq1\}$ denote the closed Bloch ball. The reduction begins with the canonical trace--Bloch form of a qubit POVM.

\begin{lemma}[Canonical Bloch disintegration]
\label{lem:povm-disintegration}
Let $M$ be a POVM on a standard Borel outcome space $\Omega$.  There is a
unique probability measure $\nu$, together with a $\nu$-almost-everywhere
unique measurable function $\br:\Omega\to\Bbar$, such that
\begin{equation}
 M(\dd\omega)
 =
 \left(\I+\br(\omega)\cdot\bsigma\right)\nu(\dd\omega),
 \qquad
 \int_{\Omega}\br(\omega)\nu(\dd\omega)=0.
 \label{eq:povm-disintegration}
\end{equation}
The density is rank one exactly when $\|\br(\omega)\|=1$, and rank two
exactly when $\|\br(\omega)\|<1$.
\end{lemma}

For each conditional local POVM, let
\begin{equation}
 \mu_{\lambda j}
 =
 (\br_{\lambda j})_{\#}\nu_{\lambda j}
 \in\calP(\Bbar)
 \label{eq:pushforward-measure}
\end{equation}
be its trace--Bloch pushforward, where $\calP(\Bbar)$ denotes the Borel probability measures on $\Bbar$.  Define
\begin{align}
 \mu_{\calE}
 &=
 \frac1N
 \int_{\Lambda}
 \sum_{j=1}^{N}\mu_{\lambda j}\,
 \xi(\dd\lambda),
 \label{eq:aggregate-measure}\\
 \calZ
 &=
 \left\{
  \mu\in\calP(\Bbar):
  \int_{\Bbar}\br\,\mu(\dd\br)=0
 \right\}.
 \label{eq:design-space}
\end{align}

\begin{theorem}[Fixed-readout reduction]
\label{thm:architecture}
Every fixed-local architecture has Fisher operator
\begin{equation}
 F_{\calE}(\bn)
 =
 N\eta^2P_{\bn}
 \left[
  \int_{\Bbar}
  \frac{\br\br^T}{1+\eta\br\cdot\bn}
  \,\mu_{\calE}(\dd\br)
 \right]
 P_{\bn}.
 \label{eq:aggregate-fisher}
\end{equation}
Conversely, every $\mu\in\calZ$ is realized by the parameter-independent
qubit POVM
\begin{equation}
 M_{\mu}(A)
 =
 \int_A
 \left(\I+\br\cdot\bsigma\right)\mu(\dd\br),
 \qquad A\subseteq\Bbar,
 \label{eq:converse-povm}
\end{equation}
used at every node.  Optimization over all fixed-local architectures is
therefore exactly optimization over $\calZ$.

Let
\begin{equation}
 B_{\mu}=\Span(\supp\mu)\subseteq\R^3.
 \label{eq:bloch-span}
\end{equation}
Then, for every $\bn\in S^2$,
\begin{align}
 \ran F_{\mu}(\bn)&=P_{\bn}B_{\mu},
 \label{eq:fisher-support}\\
 \rank F_{\mu}(\bn)&=\dim(P_{\bn}B_{\mu}).
 \label{eq:fisher-rank}
\end{align}
The Fisher field has rank two at every direction, and the complete recorded
experiment is injective on $S^2$, if and only if $B_{\mu}=\R^3$.
\end{theorem}

Thus all detector choices and recorded randomizations enter the minimax
problem only through one aggregate design measure.  Global regularity is a
span condition, not an outcome-count condition.  The corresponding finite
detector-degree law and its sharp realizations are stated and proved in
Appendix~\ref{app:reduction}.

\section{Local benchmark and Haar minimax rigidity}
\label{sec:minimax}

If the operating direction is known and local measurements may be retuned,
the trace bound in Appendix~\ref{app:local} gives, for homogeneous probes,
\begin{equation}
 \Tr[F(\bn)^{-1}]\geq\frac{4}{N\eta^2}.
 \label{eq:pointwise-homogeneous}
\end{equation}
Equality is attained by a tight frame of rank-one projective directions in
the tangent plane.  This pointwise value is the natural benchmark for a
single readout that must work for all $\bn$.  Deterministic continuous
calibration can have a topological obstruction, whereas recorded
randomization always attains the pointwise value; the exact weighted-frame
and Euler-class statements are collected in Appendix~\ref{app:local}.

Let $\omega$ be normalized Haar surface measure on $S^2$, and define
\begin{align}
 f(\eta)
 &=
 \frac{\eta^2}{4}
 \int_{-1}^{1}
 \frac{1-t^2}{1+\eta t}\,\dd t,
 \label{eq:f-integral}\\
 &=
 \frac{
  2\eta-(1-\eta^2)
  \log\!\left(\frac{1+\eta}{1-\eta}\right)
 }{4\eta}.
 \label{eq:f-closed}
\end{align}
The spin-coherent covariant POVM is
\begin{equation}
 M_{\mathrm H}(\dd\bu)
 =
 \left(\I+\bu\cdot\bsigma\right)\omega(\dd\bu),
 \qquad \bu\in S^2,
 \label{eq:Haar-povm}
\end{equation}
and its one-copy Fisher operator is
\begin{equation}
 F_{\mathrm H}^{(1)}(\bn)=f(\eta)P_{\bn}.
 \label{eq:Haar-fisher}
\end{equation}

\begin{theorem}[Haar minimax rigidity]
\label{thm:Haar-rigidity}
For every trace-balanced spectral loss and every $0<\eta<1$,
\begin{equation}
 \inf_{\calE\,\mathrm{fixed\ local}}
 \sup_{\bn\in S^2}
 L_{\ell}(F_{\calE}(\bn))
 =
 \ell\!\left(Nf(\eta),Nf(\eta)\right).
 \label{eq:general-minimax-value}
\end{equation}
A fixed-local architecture attains this value if and only if
\begin{equation}
 \mu_{\calE}=\omega
 \quad\text{as a probability measure on }S^2\subset\Bbar.
 \label{eq:Haar-uniqueness}
\end{equation}
Every positively weighted constituent is therefore rank one almost everywhere,
and the aggregate directional distribution is Haar.  For $A$ loss,
\begin{equation}
 C_{\mathrm{fix,loc}}^{\star}(N,\eta)
 =
 \frac{2}{Nf(\eta)}.
 \label{eq:A-minimax-value}
\end{equation}
If every node uses the same fixed POVM, the Haar POVM is unique up to null
sets, outcome refinements, and measurable relabeling.
\end{theorem}

The distinction between optimality and rigidity is essential.  Rotational
symmetrization constructs the Haar competitor, but uniqueness follows only
from the equality conditions.  The rotationally averaged Fisher trace is at
most $2Nf(\eta)$, with equality only for rank-one designs.  Minimax equality
then forces
\begin{equation}
 F_{\mu}(\bn)=Nf(\eta)P_{\bn}
 \quad\text{for every }\bn.
 \label{eq:isotropic-Fisher-field}
\end{equation}
Taking the trace gives a spherical convolution with kernel
\begin{equation}
 g_{\eta}(t)=\frac{1-t^2}{1+\eta t}.
 \label{eq:g-kernel}
\end{equation}
Its Legendre coefficient is nonzero in every degree $l\geq2$.  Mass and
barycenter remove degrees zero and one, so the design measure must be Haar.
The complete proof is given in Appendix~\ref{app:minimax}.

The exact cost of forbidding pointwise retuning is therefore
\begin{equation}
 \Gamma_{\mathrm{fix}}(\eta)
 =
 \frac{C_{\mathrm{fix,loc}}^{\star}}
 {4/(N\eta^2)}
 =
 \frac{\eta^2}{2f(\eta)}.
 \label{eq:global-penalty}
\end{equation}
It decreases strictly from $3/2$ as $\eta\to0^+$ to $1$ as
$\eta\to1^-$.  The endpoint value does not follow by regular continuation:
at $\eta=1$, zero-probability outcomes alter the Fisher model and the scalar
kernel loses all harmonics above degree one.

\section{Finite readouts}
\label{sec:finite}

The rigidity theorem has an immediate operational consequence before any
finite-support optimization is performed.  Since equality in the unrestricted
fixed-local problem requires the aggregate design measure itself to be Haar,
no finite-support readout can attain the mixed-state minimax value.  Finite
readout is therefore a genuinely stricter resource class, rather than merely
a discrete representation of an unrestricted optimum.  We now quantify this
separation for arbitrary support size and solve it exactly at the smallest
support compatible with global regularity.

Let $V_M^{(\ell)}(N,\eta)$ be the minimax value in
Eq.~\eqref{eq:general-minimax-value} when the aggregate Bloch measure has at
most $M$ atoms.  Repeated labels at the same Bloch vector count as one atom.
Define
\begin{equation}
 T_M(N,\eta)
 =
 N\eta^2\frac{M-2}{M-1+\eta}.
 \label{eq:TM-def}
\end{equation}

\begin{theorem}[Finite-support separation and lower bound]
\label{thm:finite-general}
For every trace-balanced loss and $0<\eta<1$,
\begin{align}
 V_M^{(\ell)}(N,\eta)&=+\infty,
 &&M\leq3,
 \label{eq:finite-M-singular}\\
 V_M^{(\ell)}(N,\eta)
 &>
 \ell\!\left(Nf(\eta),Nf(\eta)\right),
 &&4\leq M<\infty,
 \label{eq:finite-M-gap}\\
 \lim_{M\to\infty}V_M^{(\ell)}(N,\eta)
 &=
 \ell\!\left(Nf(\eta),Nf(\eta)\right).
 \label{eq:finite-M-limit}
\end{align}
For every finite $M\geq4$ the minimum exists, and
\begin{equation}
 V_M^{(\ell)}(N,\eta)
 \geq
 \max\left\{
  \ell\!\left(Nf(\eta),Nf(\eta)\right),
  \psi_{\ell}\!\left(T_M(N,\eta)\right)
 \right\}.
 \label{eq:finite-general-lower}
\end{equation}
In particular,
\begin{equation}
 V_M^{(A)}(N,\eta)
 \geq
 \max\left\{
  \frac{2}{Nf(\eta)},
  \frac{4(M-1+\eta)}{N\eta^2(M-2)}
 \right\}.
 \label{eq:finite-A-lower}
\end{equation}
For each fixed $0<\eta<1$, there are constants
$c_{\eta},C_{\eta}>0$ such that, for all sufficiently large $M$,
\begin{equation}
 0<
 V_M^{(A)}(N,\eta)-\frac{2}{Nf(\eta)}
 \leq
 \frac{C_{\eta}}{N}\ee^{-c_{\eta}\sqrt M}.
 \label{eq:finite-rate}
\end{equation}
Thus the displayed estimate is a stretched-exponential upper convergence
bound; no matching lower rate is asserted.
\end{theorem}

\begin{theorem}[Exact four-atom minimax design]
\label{thm:tetrahedron}
For every $0<\eta<1$,
\begin{align}
 V_4^{(A)}(N,\eta)
 &=
 \frac{2(3+\eta)}{N\eta^2},
 \label{eq:tetra-A-value}\\
 V_4^{(D)}(N,\eta)
 &=
 \frac{3+\eta}{N\eta^2}.
 \label{eq:tetra-D-value}
\end{align}
For either loss, the unique minimizing aggregate design, up to a rotation and
permutation of the atoms, is
\begin{equation}
 \mu_{\mathrm{tet}}
 =
 \frac14\sum_{a=1}^{4}\delta_{\bu_a},
 \qquad
 \bu_a\cdot\bu_b=-\frac13
 \quad(a\neq b),
 \label{eq:tetra-design}
\end{equation}
equivalently the tetrahedral qubit SIC
\begin{equation}
 E_a=\frac14(\I+\bu_a\cdot\bsigma).
 \label{eq:tetra-effects}
\end{equation}
Its worst directions are exactly
\begin{equation}
 \bn=-\bu_a,
 \qquad a=1,\ldots,4,
 \label{eq:tetra-worst}
\end{equation}
and at each such direction
\begin{equation}
 F_{\mathrm{tet}}(-\bu_a)
 =
 \frac{N\eta^2}{3+\eta}P_{-\bu_a}.
 \label{eq:tetra-worst-F}
\end{equation}
\end{theorem}

The general bound is obtained by promoting every atom to the unit sphere and
evaluating the Fisher trace at the antipode of a largest-weight atom.  At
$M=4$, equality forces four equal unit weights; strict Jensen equality then
forces all pairwise inner products to be $-1/3$.  The remaining task is not a
symmetry assumption but a global verification that no other direction is
worse.  Appendix~\ref{app:finite} gives the exact rational trace and
determinant fields and proves that equality occurs only at the four
antipodes.

\section{Robustness of Haar rigidity}
\label{sec:robustness}

The preceding proof uses both a scalar trace field and a traceless
rank-two tensor field.  Unequal contrasts separate these two constraints,
while the pure endpoint removes the higher harmonics of the scalar one.  The
resulting classifications show precisely which part of Haar rigidity
survives.

\subsection{Unequal contrasts}
\label{subsec:heterogeneous}

Let node $j$ have contrast $0<\eta_j<1$, setting-averaged Bloch measure
$\mu_j\in\calZ$, and put
\begin{equation}
 A=\sum_{j=1}^{N}f(\eta_j).
 \label{eq:heterogeneous-A}
\end{equation}
Let $\alpha_1,\ldots,\alpha_K$ be the distinct contrast values, define
\begin{equation}
 G_a=\{j:\eta_j=\alpha_a\},
 \qquad
 \Delta_a=\sum_{j\in G_a}(\mu_j-\omega),
 \label{eq:contrast-groups}
\end{equation}
and, for an orthonormal spherical-harmonic basis $\{Y_{lm}\}$, write
\begin{equation}
 \widehat\Delta_{a,lm}
 =
 \int_{S^2}\overline{Y_{lm}(\bu)}\,\Delta_a(\dd\bu).
 \label{eq:Delta-harmonics}
\end{equation}
Let $P_l$ denote the degree-$l$ Legendre polynomial.  For $l\geq2$, define
\begin{align}
 a_l(\alpha)
 &=
 \alpha^2
 \int_{-1}^{1}
 \frac{1-t^2}{1+\alpha t}P_l(t)\,\dd t,
 \label{eq:a-multiplier}\\
 b_l(\alpha)
 &=
 \int_{-1}^{1}
 (1+\alpha t)\log(1+\alpha t)P_l(t)\,\dd t.
 \label{eq:b-multiplier}
\end{align}

\begin{theorem}[Heterogeneous equality classification]
\label{thm:heterogeneous-fixed}
For every trace-balanced loss,
\begin{equation}
 \inf_{\calE\,\mathrm{fixed\ local}}
 \sup_{\bn\in S^2}L_{\ell}(F_{\calE}(\bn))
 =
 \ell(A,A).
 \label{eq:heterogeneous-minimax}
\end{equation}
For $A$ loss, $C_{\mathrm{fix,loc}}^{\star}=2/A$.

An architecture attains Eq.~\eqref{eq:heterogeneous-minimax} if and only if
every $\mu_j$ is supported on $S^2$ and, for all $l\geq2$ and
$-l\leq m\leq l$,
\begin{align}
 \sum_{a=1}^{K}a_l(\alpha_a)\widehat\Delta_{a,lm}&=0,
 \label{eq:hetero-trace-condition}\\
 \sum_{a=1}^{K}b_l(\alpha_a)\widehat\Delta_{a,lm}&=0.
 \label{eq:hetero-tensor-condition}
\end{align}
For each $l\geq2$, the ratio
\begin{equation}
 R_l(\alpha)=\frac{a_l(\alpha)}{b_l(\alpha)}
 \label{eq:multiplier-ratio}
\end{equation}
is strictly increasing on $(0,1)$.  Consequently:
\begin{enumerate}
 \item if $K=1$, the total aggregate is Haar,
 \begin{equation}
  \sum_{j=1}^{N}\mu_j=N\omega;
  \label{eq:one-contrast-equality}
 \end{equation}
 \item if $K=2$, each contrast-group aggregate is separately Haar,
 \begin{equation}
  \sum_{j\in G_a}\mu_j=|G_a|\omega,
  \qquad a=1,2;
  \label{eq:two-contrast-equality}
 \end{equation}
 \item if $K\geq3$, non-Haar group aggregates satisfying
 Eqs.~\eqref{eq:hetero-trace-condition} and
 \eqref{eq:hetero-tensor-condition} exist.
\end{enumerate}
\end{theorem}

The transition occurs because isotropy imposes two independent linear
conditions on each harmonic mode.  Two distinct contrast classes give an
invertible $2\times2$ multiplier matrix; a third class creates a null
direction and permits exact anisotropic cancellation.  The statement concerns
each group aggregate, not each individual node measure.  The multiplier
analysis and construction are given in Appendix~\ref{app:heterogeneous}.

The pointwise-to-fixed $A$-loss ratio is
\begin{equation}
 \Gamma_{\mathrm{fix}}(\eta_1,\ldots,\eta_N)
 =
 \frac{\sum_j\eta_j^2}{2\sum_jf(\eta_j)},
 \qquad
 1<\Gamma_{\mathrm{fix}}<\frac32.
 \label{eq:global-penalty-heterogeneous}
\end{equation}

\subsection{Pure-state endpoint}
\label{subsec:pure}

Set $\eta=1$.  We use the ordinary nonregular convention in which
Eq.~\eqref{eq:fisher-def} is integrated only over outcomes of positive
probability \cite{ZhuHayashi2018}.  A rank-one effect exactly antipodal to
$\bn$ has zero probability and zero first derivative and contributes zero.
For a design measure $\mu\in\calZ$,
\begin{equation}
 F_{\mu}^{\mathrm{pure}}(\bn)
 =
 NP_{\bn}
 \left[
  \int_{1+\br\cdot\bn>0}
  \frac{\br\br^T}{1+\br\cdot\bn}
  \,\mu(\dd\br)
 \right]
 P_{\bn}.
 \label{eq:pure-fisher}
\end{equation}
Let $V_M^{(\ell)}(N,1)$ denote the corresponding finite-support minimax
value.

\begin{theorem}[Pure-state minimax rigidity]
\label{thm:pure}
For every trace-balanced loss,
\begin{equation}
 \inf_{\calE\,\mathrm{fixed\ local}}
 \sup_{\bn\in S^2}L_{\ell}(F_{\calE}^{\mathrm{pure}}(\bn))
 =
 \ell\!\left(\frac N2,\frac N2\right).
 \label{eq:pure-minimax}
\end{equation}
Equality holds if and only if the aggregate Bloch measure is Haar.  Hence
\begin{equation}
 C_{\mathrm{fix,loc}}^{\star}(N,1)=\frac4N,
 \label{eq:pure-A}
\end{equation}
and no finite-support fixed-local measurement attains the unrestricted value.
For finite support,
\begin{align}
 V_M^{(\ell)}(N,1)&=+\infty,
 &&M\leq3,
 \label{eq:pure-small-M}\\
 V_M^{(\ell)}(N,1)
 &\geq
 \psi_{\ell}\!\left(N\frac{M-2}{M}\right),
 &&M\geq4,
 \label{eq:pure-M-general-lower}\\
 \lim_{M\to\infty}V_M^{(\ell)}(N,1)
 &=
 \ell\!\left(\frac N2,\frac N2\right).
 \label{eq:pure-M-limit-general}
\end{align}
In particular,
\begin{equation}
 V_M^{(A)}(N,1)\geq\frac{4M}{N(M-2)}.
 \label{eq:pure-M-lower}
\end{equation}
At four atoms,
\begin{align}
 V_4^{(A)}(N,1)&=\frac8N,
 \label{eq:pure-tetra-A}\\
 V_4^{(D)}(N,1)&=\frac4N.
 \label{eq:pure-tetra-D}
\end{align}
For either loss, the tetrahedral SIC is the unique minimizer up to rotations
and permutations, and its worst directions are exactly the four antipodes.
\end{theorem}

At this endpoint every positive-probability rank-one effect saturates the
single-copy Fisher-trace bound, whereas an exactly antipodal atom deletes
Fisher trace.  The scalar kernel becomes $1-t$ and cannot identify higher
harmonics.  Haar uniqueness is instead recovered from a distributional
trace-free Hessian whose multiplier remains nonzero in every degree
$l\geq2$.  Appendix~\ref{app:pure} includes the required cutoff argument at
the antipodal singularity.

\section{Measurement-architecture costs}
\label{sec:architectures}

The fixed-local result isolates the statistical cost of committing to a
parameter-independent readout before the unknown direction is learned.  We
now relax that restriction in stages to determine which part of the loss is
removed by classical feedforward and which part requires genuinely collective
quantum measurement.  This gives a resource interpretation of the fixed
minimax value rather than a separate estimation problem.

Let $C_{\mathsf X}(N,\eta)$ be the $A$-minimax value for architecture class
$\mathsf X$, where
\begin{equation}
 \mathsf{FL}\subset\mathsf{1L}\subset\mathsf{L}
 \subset\mathsf{SEP}\subset\mathsf{COLL}
 \label{eq:architecture-inclusion}
\end{equation}
denote fixed local, one-way LOCC, unrestricted LOCC, separable POVMs, and
arbitrary collective POVMs on the $N$-copy block.  All protocols are
parameter independent and all branch labels are retained.

\begin{theorem}[Architecture hierarchy]
\label{thm:architecture-hierarchy}
For $0<\eta<1$ and every $N$,
\begin{align}
 C_{\mathsf{FL}}(N,\eta)
 &=
 \frac{2}{Nf(\eta)},
 \label{eq:hierarchy-FL}\\
 C_{\mathsf X}(N,\eta)
 &\geq
 \frac{4}{N\eta^2[1-(1-\eta)^N/3]},
 \quad
 \mathsf X\in\{\mathsf{1L},\mathsf{L},\mathsf{SEP}\},
 \label{eq:hierarchy-separable-lower}\\
 C_{\mathsf{COLL}}(N,\eta)
 &\geq
 \frac{2(1+\eta)}{N\eta^2}.
 \label{eq:hierarchy-collective-lower}
\end{align}
The separable lower bound is strictly larger than the pointwise local value
for every finite $N$.

A two-copy Fisher-symmetric collective measurement gives
\begin{align}
 C_{\mathsf{COLL}}(2m,\eta)
 &\leq
 \frac{4}{2m\eta^2},
 \label{eq:collective-even-upper}\\
 C_{\mathsf{COLL}}(2m+1,\eta)
 &\leq
 \frac{2}{m\eta^2+f(\eta)}.
 \label{eq:collective-odd-upper}
\end{align}
Moreover, for every fixed $0<\eta<1$ there are constants
$K_{\eta}<\infty$ and $N_{\eta}$ such that
\begin{equation}
 C_{\mathsf{COLL}}(N,\eta)
 \leq
 \frac{2(1+\eta)}{N\eta^2}
 +\frac{K_{\eta}}{N^2},
 \qquad N\geq N_{\eta}.
 \label{eq:collective-sharp-upper}
\end{equation}
Consequently,
\begin{align}
 \lim_{N\to\infty}N C_{\mathsf{1L}}(N,\eta)
 &=
 \lim_{N\to\infty}N C_{\mathsf{L}}(N,\eta)
 =
 \frac4{\eta^2},
 \label{eq:LOCC-asymptotic}\\
 \lim_{N\to\infty}N C_{\mathsf{SEP}}(N,\eta)
 &=
 \frac4{\eta^2},
 \label{eq:SEP-asymptotic}\\
 \lim_{N\to\infty}N C_{\mathsf{COLL}}(N,\eta)
 &=
 \frac{2(1+\eta)}{\eta^2}.
 \label{eq:COLL-asymptotic}
\end{align}
At $\eta=1$, all five classes have the exact value
\begin{equation}
 C_{\mathsf X}(N,1)=\frac4N.
 \label{eq:pure-hierarchy-collapse}
\end{equation}
\end{theorem}

The separable lower bound follows from an average trace inequality for every
rank-one product refinement.  A two-stage one-way protocol first localizes
$\bn$ on a sublinear pilot sample and then measures randomized tangent
projective lines, attaining $4/\eta^2$.  For collective readout, the lower
bound is the finite-copy Holevo bound.  Its matching upper bound is supplied
by a concrete parameter-independent POVM: first resolve the Schur spin
sector, then perform the spin-coherent covariant POVM within that sector.  Its
ordinary Fisher field is exactly isotropic and has coefficient
\begin{equation}
 \kappa_N(\eta)
 =
 \frac{N\eta^2}{1+\eta}+O_{\eta}(1),
 \label{eq:Schur-kappa-main}
\end{equation}
which gives Eq.~\eqref{eq:collective-sharp-upper}.  The construction and its
exact finite-sector Fisher formula are derived in
Appendix~\ref{app:architectures}.

For reference, the two-copy POVM consists of
\begin{align}
 M_a&=\frac34
 \bigl(|\bu_a\rangle\langle\bu_a|\bigr)^{\otimes2},
 \qquad a=1,\ldots,4,
 \label{eq:collective-SIC-effects}\\
 M_-&=\Pi_-,
 \label{eq:singlet-effect}
\end{align}
where the $|\bu_a\rangle$ form a tetrahedral SIC and $\Pi_-$ is the singlet
projector.  Its directional Fisher operator is
\begin{equation}
 F_{\mathrm{collSIC}}^{(2)}(\bn)=\eta^2P_{\bn}.
 \label{eq:collective-SIC-Fisher}
\end{equation}

\begin{table*}
\caption{\label{tab:hierarchy}
$A$-minimax comparison for homogeneous mixed probes.  The finite-$N$
separable and collective entries are lower bounds; the asymptotic
coefficients are exact.}
\begin{ruledtabular}
\begin{tabular}{lcc}
Architecture
& Finite-$N$ statement
& $\displaystyle\lim_{N\to\infty}NC(N,\eta)$
\\
\hline
Fixed local
& $\displaystyle C=\frac{2}{Nf(\eta)}$
& $\displaystyle \frac{2}{f(\eta)}$
\\[1ex]
One-way LOCC, LOCC, separable
& $\displaystyle C\geq
 \frac{4}{N\eta^2[1-(1-\eta)^N/3]}$
& $\displaystyle \frac4{\eta^2}$
\\[1ex]
Collective
& $\displaystyle C\geq\frac{2(1+\eta)}{N\eta^2}$
& $\displaystyle \frac{2(1+\eta)}{\eta^2}$
\\
\end{tabular}
\end{ruledtabular}
\end{table*}

Equation~\eqref{eq:intro-hierarchy} now has a direct resource
interpretation.  A fixed readout pays for global direction independence;
feedforward removes that cost asymptotically while retaining the local
multiparameter incompatibility; collective measurement removes the remaining
incompatibility cost.

\section{Discussion}
\label{sec:discussion}

The main structural result of this work is the rigidity of globally fixed
qubit readout.  Every parameter-independent fixed-local architecture reduces
to a zero-barycenter design measure on the Bloch ball, and a sharp rotational
trace bound determines the minimax value. Symmetry identifies the spin-coherent Haar POVM as the natural covariant
competitor; the sharp trace bound establishes its optimality, and the equality
conditions show substantially more. Optimality forces a rank-one isotropic Fisher field, and
the resulting spherical convolution equation forces the design measure itself
to be Haar.  Thus the global minimax problem has a unique aggregate solution,
not merely a convenient covariant representative.

Finite readout provides the most direct operational consequence of this
rigidity.  No finite design reaches the unrestricted mixed-state optimum.
At the smallest globally regular support, the universal atom-weight bound
forces four equal weights and regular-tetrahedral geometry, while the exact
Fisher field shows that the tetrahedral SIC is uniquely $A$- and $D$-minimax
and that its four antipodes are the only worst directions.  The heterogeneous
and pure-state analyses further identify which ingredients of the rigidity
proof are essential.  Unequal contrasts produce two independent harmonic
constraints and permit anisotropic cancellation only once three or more
contrast classes are available.  At purity the scalar harmonic mechanism
degenerates, but the traceless tensor condition remains sufficiently rigid to
recover Haar uniqueness.

The architecture hierarchy gives the corresponding resource accounting.  For
mixed states, fixed-local, adaptive/separable, and collective measurements
have three distinct asymptotic $A$-loss coefficients.  The first gap
quantifies the cost of requiring one globally fixed readout.  A two-stage
one-way LOCC protocol removes this cost asymptotically by first localizing the
direction and then measuring randomized tangent projective lines.  The
remaining multiparameter incompatibility cost is removed only by collective
measurement.  For the latter we obtain the optimal asymptotic coefficient
directly from the ordinary Fisher information of an explicit
parameter-independent Schur-coherent POVM, rather than inferring it solely
from an estimator-risk bound.

The criteria studied here are criteria for Fisher experiments: they determine
the information geometry and inverse-Fisher cost of the measurement itself,
and do not by themselves constitute a finite-sample global risk theorem for a
particular estimator.  Within that scope, the fixed-radius qubit sphere gives
an exact setting in which three resources can be separated cleanly:
state-independent global readout, adaptive classical control, and collective
quantum measurement.  The resulting distinction between symmetry and
rigidity is central: covariance identifies a natural design, whereas the
equality analysis determines when any alternative design is possible.

\appendix

\section{Measure reduction and finite regularity}
\label{app:reduction}

\subsection{Proof of Lemma~\ref{lem:povm-disintegration}}

Define the finite scalar measure
\begin{equation}
 \nu(A)=\frac12\Tr M(A).
 \label{eq:trace-measure}
\end{equation}
Because $M(A)\geq0$, $\nu(A)=0$ implies $M(A)=0$; hence $M$ is absolutely
continuous with respect to $\nu$.  The finite-dimensional Radon--Nikodym
theorem gives a positive matrix density $D(\omega)$ such that
\begin{equation}
 M(A)=\int_A D(\omega)\nu(\dd\omega).
 \label{eq:operator-RN}
\end{equation}
By definition of $\nu$, $\Tr D(\omega)=2$ for $\nu$-almost every $\omega$.
Every positive qubit operator of trace two has the unique form
\begin{equation}
 D(\omega)=\I+\br(\omega)\cdot\bsigma,
 \qquad \|\br(\omega)\|\leq1.
 \label{eq:density-Bloch}
\end{equation}
Normalization gives
\begin{equation}
 \I
 =
 M(\Omega)
 =
 \nu(\Omega)\I+
 \left(\int_{\Omega}\br\,\dd\nu\right)\cdot\bsigma.
 \label{eq:normalization-components}
\end{equation}
Taking the trace and Pauli components yields
$\nu(\Omega)=1$ and $\int\br\,\dd\nu=0$.  The eigenvalues of $D$ are
$1\pm\|\br\|$, proving the rank statements.  Uniqueness follows from the
uniqueness of the trace measure, the Radon--Nikodym density, and the Bloch
representation.

\subsection{Proof of Theorem~\ref{thm:architecture}}

For one local POVM in canonical form,
\begin{align}
 p(\dd\omega|\bn)
 &=
 \Tr[\rho_{\bn}M(\dd\omega)]
 =
 (1+\eta\br(\omega)\cdot\bn)\nu(\dd\omega),
 \label{eq:local-probability}\\
 \dd p(\dd\omega|\bn)[\bv]
 &=
 \eta\br(\omega)\cdot\bv\,\nu(\dd\omega).
 \label{eq:local-derivative}
\end{align}
Therefore
\begin{equation}
 F_M(\bn)
 =
 \eta^2P_{\bn}
 \left[
  \int_{\Bbar}
  \frac{\br\br^T}{1+\eta\br\cdot\bn}
  \,\mu(\dd\br)
 \right]P_{\bn}.
 \label{eq:one-measure-FI}
\end{equation}
Conditional independence makes local Fisher operators additive.  Recording
$\lambda$ makes Fisher information affine in the setting law, proving
Eq.~\eqref{eq:aggregate-fisher}.

Conversely, for $\mu\in\calZ$, the operator in
Eq.~\eqref{eq:converse-povm} is positive because $\|\br\|\leq1$, and
\begin{equation}
 M_{\mu}(\Bbar)
 =
 \I+
 \left(\int\br\,\dd\mu\right)\cdot\bsigma
 =
 \I.
 \label{eq:converse-normalization}
\end{equation}
It is therefore a POVM whose trace--Bloch measure is $\mu$.

For $\bv\in T_{\bn}S^2$,
\begin{equation}
 \bv^TF_{\mu}(\bn)\bv
 =
 N\eta^2
 \int_{\Bbar}
 \frac{(\br\cdot\bv)^2}{1+\eta\br\cdot\bn}
 \,\mu(\dd\br).
 \label{eq:FI-quadratic-form}
\end{equation}
The denominator is bounded below by $1-\eta>0$.  The quadratic form vanishes
if and only if $\br\cdot\bv=0$ for $\mu$-almost every $\br$, so
\begin{equation}
 \ker F_{\mu}(\bn)
 =
 T_{\bn}S^2\cap B_{\mu}^{\perp}.
 \label{eq:FI-kernel}
\end{equation}
Taking the orthogonal complement inside $T_{\bn}S^2$ gives
Eqs.~\eqref{eq:fisher-support} and \eqref{eq:fisher-rank}.

If $B_{\mu}=\R^3$, then $P_{\bn}B_{\mu}=T_{\bn}S^2$ for every $\bn$.  If
$\dim B_{\mu}=2$, choose a unit $\bn\in B_{\mu}$; then
$\dim(P_{\bn}B_{\mu})=1$.  If $\dim B_{\mu}\leq1$, rank two is impossible.
This proves the regularity equivalence.

Equality of the complete recorded laws at $\bn$ and $\bn'$ implies, after
marginalizing over every positively weighted setting and node,
\begin{equation}
 \br\cdot(\bn-\bn')=0
 \quad\text{for }\mu_{\calE}\text{-almost every }\br.
 \label{eq:injectivity-linear}
\end{equation}
If $B_{\mu}=\R^3$, then $\bn=\bn'$.  Conversely, if
$B_{\mu}\neq\R^3$, choose a unit $\bw\in B_{\mu}^{\perp}$ and a direction
with $\bn\cdot\bw\neq0$.  Its reflection
\begin{equation}
 \bn'=\bn-2(\bn\cdot\bw)\bw
 \label{eq:reflection-counterexample}
\end{equation}
is a distinct unit vector with the same inner product against every
$\br\in B_{\mu}$, so the complete probability laws coincide.

\subsection{Finite detector-degree law}

For a finite setting library, let setting $k$ have positive sampling
probability $\pi_k$, and let local slot $(k,j)$ have $m_{kj}$ nonzero outcomes
with Bloch span $B_{kj}$ and degree $d_{kj}=\dim B_{kj}$.

\begin{corollary}[Finite detector-degree law]
\label{cor:resource-law}
A finite pooled local architecture is uniformly regular and globally
injective if and only if
\begin{equation}
 \sum_{k:\,\pi_k>0}\sum_{j=1}^{N}B_{kj}=\R^3.
 \label{eq:subspace-sum-criterion}
\end{equation}
Moreover,
\begin{equation}
 3
 \leq
 \sum_{k:\,\pi_k>0}\sum_{j=1}^{N}d_{kj}
 \leq
 \sum_{k:\,\pi_k>0}\sum_{j=1}^{N}(m_{kj}-1).
 \label{eq:detector-degree-law}
\end{equation}
The lower bound is attainable.  If the rightmost sum equals three,
regularity holds exactly when every local bound is saturated and the nonzero
subspaces form a direct-sum decomposition of $\R^3$.  The irreducible degree
partitions are
\begin{equation}
 3,\qquad 2+1,\qquad 1+1+1,
 \label{eq:degree-partitions}
\end{equation}
realized respectively by a four-outcome informationally complete POVM, a
planar trine plus one transverse binary POVM, and three independent binary
projective slots.

If each of $K$ positively weighted settings applies one binary projective
axis at each of $N$ nodes, then
\begin{equation}
 K_{\min}^{\mathrm{pool}}(N)
 =
 \left\lceil\frac3N\right\rceil.
 \label{eq:pooled-threshold}
\end{equation}
\end{corollary}

For a finite local POVM, canonical normalization gives a positive relation
\begin{equation}
 \sum_{x=1}^{m_{kj}}\nu_{kjx}\br_{kjx}=0,
 \qquad \nu_{kjx}>0.
 \label{eq:positive-relation}
\end{equation}
Thus its $m_{kj}$ Bloch vectors span at most $m_{kj}-1$ dimensions.  Because
all aggregate coefficients are positive,
\begin{equation}
 B_{\mu_{\calE}}
 =
 \sum_{k:\pi_k>0}\sum_jB_{kj}.
 \label{eq:aggregate-subspace-sum}
\end{equation}
Theorem~\ref{thm:architecture} proves
Eq.~\eqref{eq:subspace-sum-criterion}; the dimension inequalities follow.
If the total available degree is three and the aggregate span is
three-dimensional, equality holds in every intermediate inequality.  Hence
each $d_{kj}=m_{kj}-1$ and the subspaces form a direct sum.  The converse is
immediate.

A tetrahedral SIC has degree three \cite{Renes2004,Scott2006}; a trine has
degree two; and a binary projective measurement has degree one.  A binary
projective slot contributes one line.  Uniform regularity therefore requires
at least three independent slots and is attained by any three spanning axes.
Since $NK$ slots are available, $NK\geq3$ is necessary and sufficient,
proving Eq.~\eqref{eq:pooled-threshold}.

\section{Pointwise benchmark, weighted frames, and calibration topology}
\label{app:local}

Let the local contrasts be $0<\eta_j<1$, and write
\begin{align}
 \rho_{\bn}^{(j)}
 &=
 \frac12(\I+\eta_j\bn\cdot\bsigma),
 \label{eq:heterogeneous-state}\\
 w_j&=\eta_j^2,
 \qquad
 S=\sum_{j=1}^{N}w_j.
 \label{eq:heterogeneous-model}
\end{align}

\begin{theorem}[Pointwise local optimum]
\label{thm:local-optimum}
For arbitrary local POVMs at a known $\bn$,
\begin{equation}
 \Tr[F(\bn)^{-1}]\geq\frac4S.
 \label{eq:local-A-bound}
\end{equation}
Equality holds if and only if, for every node $j$, the canonical Bloch vectors
of all nonzero effects are rank one and tangent at $\bn$, and their total
second moment is tight:
\begin{equation}
 \sum_{j=1}^{N}w_j
 \int_{S^2\cap T_{\bn}S^2}
 \br\br^T\,\mu_j(\dd\br)
 =
 \frac S2P_{\bn}.
 \label{eq:general-tight-equality}
\end{equation}
The bound is always attainable by independently choosing a Haar-random
projective line in $T_{\bn}S^2$ at every node and retaining the chosen line.

If node $j$ is restricted to one deterministic binary projective line
$\ell_j\subset T_{\bn}S^2$, and $\Pi_{\ell_j}$ denotes the orthogonal
projector onto that line, equality is equivalent to
\begin{equation}
 \sum_{j=1}^{N}w_j\Pi_{\ell_j}
 =
 \frac S2I_{T_{\bn}S^2}.
 \label{eq:weighted-tight-frame}
\end{equation}
Such a deterministic optimum exists if and only if
\begin{equation}
 \max_jw_j\leq\frac S2.
 \label{eq:polygon-condition}
\end{equation}
Modulo a tangent orthonormal frame, all deterministic equality configurations
are exactly the solutions of
\begin{equation}
 \sum_{j=1}^{N}w_j\ee^{2\ii\theta_j}=0,
 \qquad
 \ell_j=\Span\{(\cos\theta_j,\sin\theta_j)\}.
 \label{eq:doubled-angle-polygon}
\end{equation}
For two deterministic binary nodes, the exact optimum is
\begin{equation}
 \frac1{w_1}+\frac1{w_2},
 \label{eq:heterogeneous-deterministic-cost}
\end{equation}
attained by orthogonal tangent lines; it equals $4/(w_1+w_2)$ only when
$w_1=w_2$.
\end{theorem}

For a node of contrast $\eta_j$, put $c=\br\cdot\bn$.  Taking the tangent
trace of Eq.~\eqref{eq:one-measure-FI} gives
\begin{equation}
 \Tr_{T_{\bn}S^2}F_j(\bn)
 =
 \eta_j^2
 \int
 \frac{\|\br\|^2-c^2}{1+\eta_jc}
 \,\mu_j(\dd\br).
 \label{eq:tangent-trace-general}
\end{equation}
For $\|\br\|\leq1$ and $-1\leq c\leq1$,
\begin{align}
 \frac{\|\br\|^2-c^2}{1+\eta_jc}
 &\leq
 \frac{1-c^2}{1+\eta_jc},
 \label{eq:rank-one-step}\\
 &\leq
 1-\eta_jc,
 \label{eq:tangent-step}
\end{align}
because
\begin{equation}
 (1-\eta_jc)(1+\eta_jc)-(1-c^2)
 =
 (1-\eta_j^2)c^2\geq0.
 \label{eq:scalar-difference}
\end{equation}
Using $\int\br\,\dd\mu_j=0$ yields
\begin{equation}
 \Tr F_j(\bn)\leq\eta_j^2=w_j.
 \label{eq:single-node-trace-bound}
\end{equation}
Since $0<\eta_j<1$, equality requires $\|\br\|=1$ and $c=0$ for
$\mu_j$-almost every outcome.

Summing Eq.~\eqref{eq:single-node-trace-bound} gives $\Tr F\leq S$.  If
$\lambda_1,\lambda_2$ are the positive tangent eigenvalues, then
\begin{equation}
 \Tr(F^{-1})
 =
 \frac1{\lambda_1}+\frac1{\lambda_2}
 \geq
 \frac4{\lambda_1+\lambda_2}
 \geq
 \frac4S.
 \label{eq:AH-proof}
\end{equation}
Equality requires equality at every node and
$\lambda_1=\lambda_2=S/2$.  Under the rank-one tangent condition,
\begin{equation}
 F_j(\bn)
 =
 w_j\int\br\br^T\,\mu_j(\dd\br),
 \label{eq:tangent-second-moment}
\end{equation}
which proves Eq.~\eqref{eq:general-tight-equality} and its necessity and
sufficiency.

The invariant probability measure on the projective tangent circle obeys
\begin{equation}
 \int_{\mathbb P(T_{\bn}S^2)}
 \Pi_{\ell}\,\dd h_{\bn}(\ell)
 =
 \frac12P_{\bn},
 \label{eq:projective-Haar-moment}
\end{equation}
so independent Haar line randomization attains the bound.

For one deterministic line at node $j$, $F_j=w_j\Pi_{\ell_j}$.  In an
oriented tangent frame,
\begin{equation}
 \Pi_{\ell(\theta)}
 =
 \frac12I_2+
 \frac12
 \begin{pmatrix}
  \cos2\theta&\sin2\theta\\
  \sin2\theta&-\cos2\theta
 \end{pmatrix}.
 \label{eq:doubled-projector}
\end{equation}
Thus Eq.~\eqref{eq:weighted-tight-frame} is equivalent to the two real
components of Eq.~\eqref{eq:doubled-angle-polygon}.

For positive lengths $a_1,\ldots,a_k$, the attainable magnitudes of their
planar vector sum form the interval
\begin{equation}
 \left[
  \max\left\{0,2\max_j a_j-\sum_j a_j\right\},
  \sum_j a_j
 \right].
 \label{eq:resultant-interval}
\end{equation}
This follows inductively because adding a vector of length $a$ to a resultant
of length $r$ produces every magnitude in $[|r-a|,r+a]$.  Zero is attainable
if and only if $2\max_jw_j\leq S$, proving the polygon criterion and the
weighted-tight-frame characterization \cite{Waldron2018}.

For two general laboratory axes $\bmom_j$, let
$c_j=\bmom_j\cdot\bn$ and let $\bu_j$ be the normalized nonzero tangent
projection.  The binary Fisher contribution is
\begin{equation}
 F_{\bmom_j}^{(j)}(\bn)
 =
 \lambda_j\bu_j\bu_j^T,
 \qquad
 \lambda_j=
 \frac{w_j(1-c_j^2)}{1-w_jc_j^2}
 \leq w_j,
 \label{eq:binary-strength-general}
\end{equation}
with equality exactly for a tangent axis.  Increasing either rank-one
strength decreases the inverse-trace cost.  For tangent unit vectors
subtending angle $\alpha$,
\begin{equation}
 \Tr\!\left[
  (w_1\bu\bu^T+w_2\bv\bv^T)^{-1}
 \right]
 =
 \frac{w_1^{-1}+w_2^{-1}}{\sin^2\alpha},
 \label{eq:two-vector-inverse}
\end{equation}
which is minimized at $\alpha=\pi/2$.

The deterministic equality set has a global calibration problem.  Let $X$ be
paracompact and have the homotopy type of a CW complex, and let $E\to X$ be
an oriented Euclidean rank-two bundle.  In the sensing application,
$E=\bn_X^*(TS^2)$ for a continuous direction map
$\bn_X:X\to S^2$.  Let $\calT_w(E_x)$ be the set of labeled projective tight
frames in $E_x$ satisfying Eq.~\eqref{eq:weighted-tight-frame}.  For a fixed
frame shape $\mathsf C$, possibly modulo a weight-preserving node-permutation
group, let its orientation-preserving rotational stabilizer be $C_q$.

\begin{theorem}[Euler-class calibration and convexification]
\label{thm:euler-calibration}
Assume Eq.~\eqref{eq:polygon-condition}.
\begin{enumerate}
 \item A continuous labeled deterministic optimal calibration exists if and
 only if
 \begin{equation}
  2e(E)=0
  \quad\text{in }H^2(X;\mathbb Z).
  \label{eq:labeled-euler-condition}
 \end{equation}
 \item A continuous calibration constrained to the rotational orbit of
 $\mathsf C$ exists if and only if
 \begin{equation}
  q\,e(E)=0.
  \label{eq:orbit-euler-condition}
 \end{equation}
 \item If $X$ is a connected closed oriented surface and
 $E=\bn_X^*(TS^2)$, then
 \begin{equation}
  \langle e(E),[X]\rangle=2\deg(\bn_X).
  \label{eq:euler-degree}
 \end{equation}
 Hence a global deterministic calibration exists exactly when
 $\deg(\bn_X)=0$.  If the degree is nonzero, exactly two open continuous
 calibration charts are necessary and sufficient.
 \item For every oriented plane bundle and every positive weight vector,
 fiberwise Haar randomization over projective lines defines a canonical
 continuous probability-valued readout attaining $4/S$.  When
 Eq.~\eqref{eq:polygon-condition} holds, the randomization may be supported
 entirely on deterministic tight frames by Haar averaging one such frame over
 the simultaneous $\SO(2)$ action.
\end{enumerate}
\end{theorem}

Let $P=\Fr^+(E)$ be the principal $\SO(2)$ bundle of oriented orthonormal
frames.  Its Euler class is $e(E)\in H^2(X;\mathbb Z)$.  The projective line
bundle is
\begin{equation}
 \mathbb P(E)\simeq P/C_2,
 \label{eq:P-quotient-C2}
\end{equation}
where $C_2=\{1,-1\}$.  Under the group isomorphism
\begin{equation}
 \SO(2)/C_2\longrightarrow\SO(2),
 \qquad zC_2\longmapsto z^2,
 \label{eq:square-map}
\end{equation}
transition functions are squared.  Therefore
\begin{equation}
 e(\mathbb P(E))=2e(E).
 \label{eq:projective-euler}
\end{equation}
A principal circle bundle has a section if and only if its Euler class
vanishes \cite{MilnorStasheff1974,Husemoller1994}.

A labeled tight-frame calibration contains its first line and hence gives a
section of $\mathbb P(E)$, proving necessity of
Eq.~\eqref{eq:labeled-euler-condition}.  Conversely, a section of
$\mathbb P(E)$ reduces the structure group from $\SO(2)$ to $C_2$.  Choose
any reference weighted projective tight frame in $\R^2$ and rotate it so that
its first line is the reference axis.  The simultaneous half turn fixes every
projective line, so the reference configuration is $C_2$ invariant and is
transported by the reduced bundle to a global labeled calibration.

For a fixed physical shape $\mathsf C$, its rotational orbit is
$\SO(2)/C_q$, and the calibration bundle is $P/C_q$.  The isomorphism
$zC_q\mapsto z^q$ raises transition functions to their $q$th powers, so
\begin{equation}
 e(P/C_q)=q\,e(E).
 \label{eq:q-euler}
\end{equation}
This proves item 2.

Naturality gives $e(E)=\bn_X^*e(TS^2)$.  Hence, on a closed oriented surface,
\begin{align}
 \langle e(E),[X]\rangle
 &=
 \langle e(TS^2),(\bn_X)_*[X]\rangle
 \nonumber\\
 &=
 \deg(\bn_X)\langle e(TS^2),[S^2]\rangle
 =
 2\deg(\bn_X).
 \label{eq:euler-degree-proof}
\end{align}
Since $H^2(X;\mathbb Z)\simeq\mathbb Z$ is torsion free, every condition
$q e(E)=0$ reduces to $\deg(\bn_X)=0$.

If $e(E)=0$, the oriented plane bundle is trivial and one chart suffices.  If
$e(E)\neq0$, one chart is impossible.  Choose a point $p\in X$.  A disk
$V$ around $p$ is contractible, while $U=X\setminus\{p\}$ has the homotopy
type of a one-dimensional CW complex.  Thus
$H^2(U;\mathbb Z)=H^2(V;\mathbb Z)=0$, and $E$ is trivial over both open
sets.  Two charts suffice.

Finally, each fiber $\mathbb P(E_x)$ carries a unique invariant probability
measure $h_x$, independent of a local frame.  Equation
\eqref{eq:projective-Haar-moment} gives
\begin{equation}
 \mathbb E[F_x]
 =
 \sum_j\frac{w_j}{2}I_{E_x}
 =
 \frac S2I_{E_x}.
 \label{eq:convexified-optimal-FI}
\end{equation}
When a deterministic tight frame exists, Haar averaging its simultaneous
$\SO(2)$ orbit gives an invariant probability measure supported entirely on
fiberwise deterministic optima.

\section{Average trace and mixed-state harmonic rigidity}
\label{app:minimax}

For $0\leq z\leq1$, define continuously
\begin{equation}
 f(z)
 =
 \frac{z^2}{4}
 \int_{-1}^{1}\frac{1-t^2}{1+zt}\,\dd t,
 \qquad f(1)=\frac12.
 \label{eq:f-general-z}
\end{equation}
Pairing $t$ and $-t$ gives
\begin{equation}
 f(z)
 =
 \frac{z^2}{2}
 \int_0^1
 \frac{1-t^2}{1-z^2t^2}\,\dd t.
 \label{eq:f-paired}
\end{equation}
For $0<z<1$,
\begin{equation}
 \frac{\partial}{\partial z}
 \frac{z^2}{1-z^2t^2}
 =
 \frac{2z}{(1-z^2t^2)^2}>0,
 \label{eq:f-monotone-integrand}
\end{equation}
so $f$ is strictly increasing.  Direct integration gives
Eq.~\eqref{eq:f-closed}.

Let
\begin{equation}
 \tau_{\mu}(\bn)=\Tr_{T_{\bn}S^2}F_{\mu}(\bn).
 \label{eq:trace-field}
\end{equation}
From Eq.~\eqref{eq:aggregate-fisher},
\begin{equation}
 \tau_{\mu}(\bn)
 =
 N\eta^2
 \int_{\Bbar}
 \frac{\|\br\|^2-(\br\cdot\bn)^2}
 {1+\eta\br\cdot\bn}
 \,\mu(\dd\br).
 \label{eq:trace-field-formula}
\end{equation}
Averaging over $\bn$ and writing $\br=q\bu$ yields
\begin{equation}
 \int_{S^2}\tau_{\mu}(\bn)\omega(\dd\bn)
 =
 2N\int_{\Bbar}f(\eta\|\br\|)\mu(\dd\br)
 \leq
 2Nf(\eta).
 \label{eq:average-trace-identity}
\end{equation}
Equality holds if and only if $\|\br\|=1$ for $\mu$-almost every $\br$.
For the Haar design, rotational invariance makes the Fisher field a multiple
of $P_{\bn}$, and a tangent component is
\begin{equation}
 \eta^2\int_{S^2}
 \frac{u_x^2}{1+\eta u_z}\,\omega(\dd\bu)
 =
 \frac{\eta^2}{4}
 \int_{-1}^{1}\frac{1-t^2}{1+\eta t}\,\dd t
 =
 f(\eta),
 \label{eq:Haar-component}
\end{equation}
which proves Eq.~\eqref{eq:Haar-fisher}.

Let
\begin{equation}
 \mathcal V_{\ell}(\mu)
 =
 \sup_{\bn\in S^2}L_{\ell}(F_{\mu}(\bn)).
 \label{eq:V-mu}
\end{equation}
If some Fisher operator is singular, the minimax lower bound is immediate.
Otherwise, with tangent eigenvalues $\lambda_1(\bn),\lambda_2(\bn)$ and
$M=\mathcal V_{\ell}(\mu)$, trace balance gives
\begin{equation}
 \psi_{\ell}(\tau_{\mu}(\bn))
 \leq
 \ell(\lambda_1(\bn),\lambda_2(\bn))
 \leq M.
 \label{eq:loss-trace-chain}
\end{equation}
If $M<\psi_{\ell}(2Nf(\eta))$, strict decrease of $\psi_{\ell}$ would imply
$\tau_{\mu}(\bn)>2Nf(\eta)$ for every $\bn$, contradicting
Eq.~\eqref{eq:average-trace-identity}.  Thus
\begin{equation}
 \mathcal V_{\ell}(\mu)
 \geq
 \psi_{\ell}(2Nf(\eta))
 =
 \ell(Nf(\eta),Nf(\eta)).
 \label{eq:minimax-lower}
\end{equation}
The Haar POVM attains equality.

Suppose equality is attained.  Equation~\eqref{eq:loss-trace-chain} implies
\begin{equation}
 \tau_{\mu}(\bn)\geq2Nf(\eta)
 \quad\text{for every }\bn.
 \label{eq:trace-pointwise-lower}
\end{equation}
The average identity forces equality everywhere and support on $S^2$.  With
trace fixed, equality in Eq.~\eqref{eq:trace-balance} is necessary at every
direction, proving Eq.~\eqref{eq:isotropic-Fisher-field}.

Taking the trace and subtracting the Haar identity gives
\begin{equation}
 \int_{S^2}g_{\eta}(\bu\cdot\bn)\,\delta(\dd\bu)=0,
 \qquad
 \delta=\mu-\omega.
 \label{eq:zero-convolution}
\end{equation}
The signed measure $\delta$ has zero mass and zero first moment.  For
$l\geq2$, polynomial division gives
\begin{equation}
 g_{\eta}(t)
 =
 -\frac{t}{\eta}+\frac1{\eta^2}
 +\frac{\eta^2-1}{\eta^2(1+\eta t)}.
 \label{eq:g-division}
\end{equation}
Rodrigues' formula and $l$ integrations by parts yield
\begin{equation}
 \int_{-1}^{1}\frac{P_l(t)}{1+\eta t}\,\dd t
 =
 \frac{(-1)^l\eta^l}{2^l}
 \int_{-1}^{1}
 \frac{(1-t^2)^l}{(1+\eta t)^{l+1}}\,\dd t.
 \label{eq:Rodrigues-integral}
\end{equation}
Therefore, for every $l\geq2$,
\begin{multline}
 \int_{-1}^{1}g_{\eta}(t)P_l(t)\,\dd t
 =
 \frac{(\eta^2-1)(-1)^l\eta^{l-2}}{2^l}
 \\
 {}
 \times
 \int_{-1}^{1}
 \frac{(1-t^2)^l}{(1+\eta t)^{l+1}}\,\dd t
 \neq0.
 \label{eq:nonzero-Legendre-spectrum}
\end{multline}

Applying the Funk--Hecke formula \cite{Muller1966} to
Eq.~\eqref{eq:zero-convolution} gives, for every spherical harmonic $Y_{lm}$,
\begin{equation}
 \left[
  2\pi\int_{-1}^{1}g_{\eta}(t)P_l(t)\,\dd t
 \right]
 \int_{S^2}\overline{Y_{lm}(\bu)}\,\delta(\dd\bu)=0.
 \label{eq:Funk-Hecke-moments}
\end{equation}
Degrees zero and one vanish by mass and barycenter; every degree $l\geq2$
vanishes because the multiplier is nonzero.  Finite linear combinations of
spherical harmonics are uniformly dense in $C(S^2)$, so $\delta=0$ by the
Riesz representation theorem.  This proves Theorem~\ref{thm:Haar-rigidity}.

Finally, Eq.~\eqref{eq:f-paired} gives
\begin{equation}
 \frac{f(\eta)}{\eta^2}
 =
 \frac12\int_0^1
 \frac{1-t^2}{1-\eta^2t^2}\,\dd t.
 \label{eq:f-ratio-integral}
\end{equation}
The integrand is strictly increasing in $\eta$, and dominated convergence
gives the limits $1/3$ and $1/2$ at $0^+$ and $1^-$, respectively.  This
proves the monotonicity and limits in Eq.~\eqref{eq:global-penalty} and the
heterogeneous bounds in Eq.~\eqref{eq:global-penalty-heterogeneous}.

\section{Finite support and the tetrahedral calculation}
\label{app:finite}

\subsection{Radial promotion and the atom-weight bound}

Let an atomic design be
\begin{equation}
 \mu=\sum_{a=1}^{m}p_a\delta_{s_a\bu_a},
 \qquad
 0\leq s_a\leq1,
 \qquad m\leq M,
 \label{eq:atomic-radial}
\end{equation}
with distinct nonzero directions after combining repetitions.  If
$c=\sum_ap_as_a>0$, define
\begin{equation}
 q_a=\frac{p_as_a}{c},
 \qquad
 \nu=\sum_aq_a\delta_{\bu_a}.
 \label{eq:radial-promotion}
\end{equation}
The zero-barycenter condition gives $\sum_aq_a\bu_a=0$.  For every
$-1\leq t\leq1$,
\begin{equation}
 \frac{s_a}{1+\eta s_at}
 \leq
 \frac1{1+\eta t},
 \label{eq:radial-inequality}
\end{equation}
because it is equivalent to $s_a\leq1$.  Consequently, in Loewner order,
\begin{equation}
 F_{\mu}(\bn)\leq cF_{\nu}(\bn)\leq F_{\nu}(\bn)
 \quad\text{for every }\bn.
 \label{eq:radial-Loewner}
\end{equation}
If $c=0$, the Fisher field is zero.  Thus radial contraction cannot increase
the smallest directional Fisher trace.

Now let $\nu=\sum_ap_a\delta_{\bu_a}$ be a zero-barycenter unit-sphere
design, and choose an atom of maximal weight $p=p_k$.  Necessarily
$p\leq1/2$, because the vector sum of all other atoms has norm at most
$1-p$ and must cancel a vector of norm $p$.  At $\bn=-\bu_k$, put
$x_a=\bu_a\cdot\bu_k$ for $a\neq k$.  Then
\begin{equation}
 \sum_{a\neq k}\frac{p_a}{1-p}x_a
 =
 -\frac{p}{1-p}.
 \label{eq:conditional-mean}
\end{equation}
The scalar function
\begin{equation}
 h_{\eta}(x)=\frac{1-x^2}{1-\eta x}
 \label{eq:h-eta}
\end{equation}
is strictly concave on $[-1,1]$ for $0<\eta<1$, since
\begin{equation}
 h_{\eta}''(x)
 =
 -\frac{2(1-\eta^2)}{(1-\eta x)^3}<0.
 \label{eq:h-concavity}
\end{equation}
Jensen's inequality and Eq.~\eqref{eq:conditional-mean} give
\begin{align}
 \frac{\tau_{\nu}(-\bu_k)}{N\eta^2}
 &=
 \sum_{a\neq k}p_a h_{\eta}(x_a)
 \nonumber\\
 &\leq
 \frac{1-2p}{1-(1-\eta)p}.
 \label{eq:atom-trace-bound}
\end{align}
The right-hand side has derivative
$-(1+\eta)/[1-(1-\eta)p]^2<0$.  Since $p\geq1/m\geq1/M$,
\begin{equation}
 \min_{\bn}\tau_{\mu}(\bn)
 \leq
 N\eta^2\frac{M-2}{M-1+\eta}
 =T_M(N,\eta).
 \label{eq:min-trace-M}
\end{equation}
At a direction realizing this trace bound,
$L_{\ell}(F)\geq\psi_{\ell}(\Tr F)$, which proves
Eq.~\eqref{eq:finite-general-lower}.

A zero-barycenter probability measure with at most three atoms has support
span of dimension at most two, proving Eq.~\eqref{eq:finite-M-singular}.  For
fixed $M$, atomic measures can be parameterized, allowing zero weights, by the
compact set
\begin{equation}
 \left\{(p_a,\br_a)_{a=1}^{M}:
 p_a\geq0,\ \sum_ap_a=1,\ \sum_ap_a\br_a=0\right\}.
 \label{eq:atomic-compact}
\end{equation}
The mixed-state Fisher kernel is uniformly continuous on
$S^2\times\Bbar$, and the extended worst-case loss is lower semicontinuous.
A minimum therefore exists for $M\geq4$.  If a finite minimizer attained the
unrestricted value, Theorem~\ref{thm:Haar-rigidity} would force a finite
measure to equal the nonatomic Haar measure.  This proves the strict gap.
Weakly convergent zero-barycenter atomic approximations to Haar prove the
limit for general trace-balanced losses.

\subsection{Equality at four atoms}

Equation~\eqref{eq:f-ratio-integral} gives
$f(\eta)/\eta^2>1/3>1/(3+\eta)$ for $0<\eta<1$.  Hence, for both $A$ and
$D$ loss at $M=4$, the atom-weight term in
Eq.~\eqref{eq:finite-general-lower} strictly dominates the unrestricted Haar
bound.  Suppose a four-atom $A$- or $D$-minimizer attains this lower bound.
Equality in Eq.~\eqref{eq:radial-Loewner} and in the trace bound forces
$c=1$, hence $s_a=1$ for every positive-weight atom.  It also forces the
maximal weight to equal $1/4$, so all four weights are $1/4$.  Repeating the
same lower-bound chain at the antipode of each now-maximal atom and applying
strict Jensen equality gives
\begin{equation}
 \bu_a\cdot\bu_b=-\frac13
 \quad(a\neq b).
 \label{eq:Jensen-tetra}
\end{equation}
Thus equality in the universal lower bound can occur only for the regular
tetrahedron.  It remains to prove that the tetrahedron never has a worse
direction.

Choose coordinates
\begin{equation}
 \begin{aligned}
 \bu_1&=(1,1,1)/\sqrt3,
 &\qquad \bu_2&=(1,-1,-1)/\sqrt3,\\
 \bu_3&=(-1,1,-1)/\sqrt3,
 &\qquad \bu_4&=(-1,-1,1)/\sqrt3.
 \end{aligned}
 \label{eq:tetra-coordinates}
\end{equation}
For $\bn=(x,y,z)$, put
\begin{equation}
 q=xyz,
 \qquad
 s=x^2y^2+y^2z^2+z^2x^2,
 \qquad
 t_a=\bu_a\cdot\bn,
 \label{eq:q-s-def}
\end{equation}
and
\begin{equation}
 \Delta=\prod_{a=1}^{4}(1+\eta t_a).
 \label{eq:Delta-def}
\end{equation}
The elementary symmetric polynomials of the $t_a$ are
\begin{equation}
 e_1=0,
 \qquad
 e_2=-\frac23,
 \qquad
 e_3=\frac{8q}{3\sqrt3},
 \qquad
 e_4=\frac{1-4s}{9}.
 \label{eq:t-symmetric}
\end{equation}
Hence
\begin{equation}
 \Delta
 =
 1-\frac23\eta^2
 +\frac{8\eta^3q}{3\sqrt3}
 +\frac{\eta^4(1-4s)}9.
 \label{eq:Delta-expanded}
\end{equation}
Write $F_{\mathrm{tet}}=N\eta^2G$.  The trace numerator over the common
denominator $\Delta$ is
\begin{align}
 4\Delta\,\Tr G
 &=
 \sum_a(1-t_a^2)\prod_{b\neq a}(1+\eta t_b)
 \nonumber\\
 &=
 4+2(1+\eta^2)e_2+(3\eta+\eta^3)e_3+4\eta^2e_4.
 \label{eq:tetra-trace-symmetric}
\end{align}
For the determinant, the Cauchy--Binet identity gives
\begin{equation}
 \det G
 =
 \frac1{16}\sum_{a<b}
 \frac{[\bn\cdot(\bu_a\times\bu_b)]^2}
 {(1+\eta t_a)(1+\eta t_b)}.
 \label{eq:Cauchy-Binet}
\end{equation}
If $\{c,d\}$ is complementary to $\{a,b\}$, then
$\bu_a\times\bu_b=\pm(\bu_c-\bu_d)/\sqrt3$.  Relabeling complementary
pairs therefore yields
\begin{align}
 16\Delta\det G
 &=
 \frac13\sum_{c<d}(t_c-t_d)^2
 (1+\eta t_c)(1+\eta t_d)
 \nonumber\\
 &=
 \frac13\left(-8e_2+12\eta e_3-4\eta^2e_2^2\right).
 \label{eq:tetra-det-symmetric}
\end{align}
Substitution of Eq.~\eqref{eq:t-symmetric} into
Eqs.~\eqref{eq:tetra-trace-symmetric} and
\eqref{eq:tetra-det-symmetric} gives
\begin{align}
 \Tr G
 &=
 \frac{2[3-\eta^2-2\eta^2s
 +\sqrt3\eta(3+\eta^2)q]}{9\Delta},
 \label{eq:tetra-trace-G}\\
 \det G
 &=
 \frac{3-\eta^2+6\sqrt3\eta q}{27\Delta}.
 \label{eq:tetra-det-G}
\end{align}

For the $A$ loss,
\begin{multline}
 \Tr[F_{\mathrm{tet}}(\bn)^{-1}]
 =
 \frac6{N\eta^2}
 \\
 {}
 \times
 \frac{3-\eta^2-2\eta^2s
 +\sqrt3\eta(3+\eta^2)q}
 {3-\eta^2+6\sqrt3\eta q}.
 \label{eq:tetra-A-field}
\end{multline}
Subtracting from the proposed worst value gives
\begin{multline}
 \frac{2(3+\eta)}{N\eta^2}
 -\Tr[F_{\mathrm{tet}}(\bn)^{-1}]
 \\
 =
 \frac{2[3-\eta^2+6\eta s
 +3\sqrt3(1+\eta)(3-\eta)q]}
 {N\eta[3-\eta^2+6\sqrt3\eta q]}.
 \label{eq:tetra-A-gap}
\end{multline}
The denominator is positive for $0<\eta<1$.  If $q\geq0$, the numerator is
positive.  If $q<0$, set
\begin{equation}
 a=3\sqrt3|q|\in(0,1].
 \label{eq:a-def}
\end{equation}
The arithmetic--geometric mean inequality gives
\begin{equation}
 s\geq3|q|^{4/3}=\frac{a^{4/3}}3,
 \label{eq:s-lower}
\end{equation}
and convexity of $a^{4/3}$ at $a=1$ gives
\begin{equation}
 a-a^{4/3}\leq\frac{1-a}{3}.
 \label{eq:a-inequality}
\end{equation}
The numerator in Eq.~\eqref{eq:tetra-A-gap} is therefore bounded below by
\begin{equation}
 (3-\eta^2)(1-a)+2\eta(a^{4/3}-a)
 \geq
 \left(3-\eta^2-\frac{2\eta}{3}\right)(1-a)>0
 \label{eq:A-gap-lower}
\end{equation}
for $a<1$.  Equality requires $a=1$ and equality in
Eq.~\eqref{eq:s-lower}, hence
$|x|=|y|=|z|=1/\sqrt3$ with $q<0$.  These are exactly the four antipodes.
At an antipode, the vanishing projected atom drops out and the remaining
three projected vectors form a tangent trine, giving
Eq.~\eqref{eq:tetra-worst-F} directly.

For the $D$ loss, Eq.~\eqref{eq:tetra-D-value} is equivalent to
\begin{equation}
 \det G\geq\frac1{(3+\eta)^2}.
 \label{eq:D-det-target}
\end{equation}
Using Eq.~\eqref{eq:Delta-expanded}, the numerator of the difference is
\begin{align}
 &\frac{(3+\eta)^2(3-\eta^2+6\sqrt3\eta q)-27\Delta}{2\eta}
 \nonumber\\
 &\quad=
 9+6\eta-3\eta^2-2\eta^3+6\eta^3s
 +9\sqrt3(3+2\eta-\eta^2)q.
 \label{eq:D-gap-expanded}
\end{align}
It is positive for $q\geq0$.  For $q<0$, Eqs.~\eqref{eq:a-def} and
\eqref{eq:s-lower} give the lower bound
\begin{equation}
 3(3+2\eta-\eta^2)(1-a)
 -2\eta^3(1-a^{4/3}).
 \label{eq:D-gap-lower}
\end{equation}
Convexity also gives
$1-a^{4/3}\leq\frac43(1-a)$, so
Eq.~\eqref{eq:D-gap-lower} is strictly positive for $a<1$.  Equality again
occurs exactly at the four antipodes.  This completes the proof of
Theorem~\ref{thm:tetrahedron}, including uniqueness.

\subsection{Stretched-exponential upper convergence bound}

Fix $0<\eta<1$ and choose
\begin{equation}
 1<\rho<\eta^{-1}+\sqrt{\eta^{-2}-1}.
 \label{eq:Bernstein-rho}
\end{equation}
The function $(1+\eta t)^{-1}$ is analytic in the Bernstein ellipse of
parameter $\rho$, so there are degree-$L$ polynomials $p_L$ satisfying
\begin{equation}
 \sup_{t\in[-1,1]}
 \left|\frac1{1+\eta t}-p_L(t)\right|
 \leq C_{\eta,\rho}\rho^{-L}.
 \label{eq:polynomial-approx}
\end{equation}
Let $\{\bu_a\}_{a=1}^{M}$ be an equal-weight spherical $(L+2)$-design.  It
integrates every matrix entry of
$\bu\bu^Tp_L(\bu\cdot\bn)$ exactly, for every fixed $\bn$.  Therefore
\begin{equation}
 \sup_{\bn\in S^2}
 \left\|F_M(\bn)-Nf(\eta)P_{\bn}\right\|
 \leq
 C'_{\eta,\rho}N\rho^{-L}.
 \label{eq:design-Fisher-rate}
\end{equation}
Spherical $t$-designs on $S^2$ exist with every sufficiently large number
$M\geq Ct^2$ of points \cite{Delsarte1977,Bondarenko2013}.  Taking
$L\asymp\sqrt M$ and using the resolvent identity on a uniformly positive
spectral interval gives Eq.~\eqref{eq:finite-rate}.

\section{Heterogeneous harmonic equality}
\label{app:heterogeneous}

For node $j$, the rotational average in
Eq.~\eqref{eq:average-trace-identity} gives
\begin{align}
 \int_{S^2}\Tr F(\bn)\,\omega(\dd\bn)
 &=
 2\sum_{j=1}^{N}
 \int_{\Bbar}f(\eta_j\|\br\|)\mu_j(\dd\br)
 \nonumber\\
 &\leq
 2\sum_{j=1}^{N}f(\eta_j)=2A.
 \label{eq:heterogeneous-average-trace}
\end{align}
The trace-balanced argument from Appendix~\ref{app:minimax} proves the lower
bound in Eq.~\eqref{eq:heterogeneous-minimax}; Haar POVMs give
$F(\bn)=AP_{\bn}$ and attain it.  Equality forces every $\mu_j$ to be
supported on $S^2$ and forces
\begin{equation}
 F(\bn)=AP_{\bn}
 \quad\text{for every }\bn.
 \label{eq:heterogeneous-isotropic-field}
\end{equation}

For a finite signed measure $\nu$ on $S^2$, define
\begin{equation}
 \mathcal F_{\alpha,\nu}(\bn)
 =
 \alpha^2P_{\bn}
 \left[
  \int_{S^2}
  \frac{\bu\bu^T}{1+\alpha\bu\cdot\bn}
  \,\nu(\dd\bu)
 \right]P_{\bn}.
 \label{eq:F-alpha-nu}
\end{equation}
Its trace has zonal kernel
$\alpha^2(1-t^2)/(1+\alpha t)$, so the Funk--Hecke multiplier in degree $l$
is $a_l(\alpha)$ up to the common nonzero factor $2\pi$.

The traceless part is encoded by a scalar potential.  Put
\begin{align}
 h_{\alpha}(t)
 &=
 (1+\alpha t)\log(1+\alpha t),
 \label{eq:hetero-kernel-potential}\\
 H_{\alpha,\nu}(\bn)
 &=
 \int_{S^2}h_{\alpha}(\bu\cdot\bn)\nu(\dd\bu).
 \label{eq:hetero-potential}
\end{align}
Since $h_{\alpha}''(t)=\alpha^2/(1+\alpha t)$ and, for a zonal function,
\begin{equation}
 \nabla^2 h(\bu\cdot\bn)
 =
 h''(t)(P_{\bn}\bu)(P_{\bn}\bu)^T
 -t h'(t)P_{\bn},
 \label{eq:zonal-Hessian}
\end{equation}
writing $\TF A=A-\tfrac12\Tr_{T_{\bn}S^2}(A)P_{\bn}$ for the trace-free part of a tangent operator, we have the exact tensor identity
\begin{equation}
 \TF\nabla^2H_{\alpha,\nu}(\bn)
 =
 \TF\mathcal F_{\alpha,\nu}(\bn).
 \label{eq:TF-Hessian-identity}
\end{equation}
On $S^2$, the kernel of the trace-free Hessian on scalar distributions is
exactly the direct sum of spherical-harmonic degrees zero and one.  Indeed,
for a degree-$l$ harmonic $Y$, the Bochner identity gives
\begin{equation}
 \int_{S^2}|\TF\nabla^2Y|^2\,\dd\omega
 =
 \frac{(l-1)l(l+1)(l+2)}2
 \int_{S^2}|Y|^2\,\dd\omega,
 \label{eq:TF-Hessian-norm}
\end{equation}
which vanishes exactly for $l=0,1$.  Thus the traceless equation in
Eq.~\eqref{eq:heterogeneous-isotropic-field} is equivalent, in degree
$l\geq2$, to the multiplier condition involving $b_l(\alpha)$.
Because every $\Delta_a$ has zero mass and zero first moment, the trace and
traceless equations are exactly Eqs.~\eqref{eq:hetero-trace-condition} and
\eqref{eq:hetero-tensor-condition}.  This proves necessity and sufficiency.

It remains to classify cancellation across distinct contrasts.  Rodrigues'
formula and $l$ integrations by parts give, for $l\geq2$,
\begin{align}
 b_l(\alpha)
 &=
 \frac{(-1)^l\alpha^l}{2^ll(l-1)}
 \int_{-1}^{1}
 \frac{(1-t^2)^l}{(1+\alpha t)^{l-1}}\,\dd t,
 \label{eq:b-closed}\\
 a_l(\alpha)
 &=
 \frac{(\alpha^2-1)(-1)^l\alpha^l}{2^l}
 \int_{-1}^{1}
 \frac{(1-t^2)^l}{(1+\alpha t)^{l+1}}\,\dd t.
 \label{eq:a-closed}
\end{align}
Both multipliers are nonzero on $(0,1)$.  The spherical Laplacian identity
\begin{equation}
 \alpha^2\frac{1-t^2}{1+\alpha t}
 =
 \Delta_{S^2}h_{\alpha}(t)
 +2\alpha\,\partial_{\alpha}h_{\alpha}(t)
 \label{eq:laplacian-alpha-identity}
\end{equation}
implies
\begin{equation}
 a_l(\alpha)
 =
 2\alpha b_l'(\alpha)-l(l+1)b_l(\alpha).
 \label{eq:a-b-relation}
\end{equation}

The absolutely convergent power series
\begin{equation}
 (1+x)\log(1+x)
 =
 x+\sum_{n=2}^{\infty}\frac{(-1)^nx^n}{n(n-1)}
 \label{eq:h-series}
\end{equation}
and the positivity of
$\int_{-1}^{1}t^{l+2k}P_l(t)\,\dd t$ show that
\begin{equation}
 b_l(\alpha)=(-1)^l\alpha^lC_l(\alpha^2),
 \label{eq:b-positive-series}
\end{equation}
where $C_l(z)=\sum_{k\geq0}c_{lk}z^k$ has strictly positive coefficients and
at least two nonzero terms.  Combining
Eqs.~\eqref{eq:a-b-relation} and \eqref{eq:b-positive-series} gives
\begin{equation}
 R_l(\alpha)
 =
 -l(l-1)+4z\frac{C_l'(z)}{C_l(z)},
 \qquad z=\alpha^2.
 \label{eq:R-series}
\end{equation}
Under the probability weights proportional to $c_{lk}z^k$, the quantity
$zC_l'(z)/C_l(z)$ is the mean of the exponent $k$.  Its derivative with
respect to $\log z$ is the variance of $k$, which is strictly positive.
Therefore $R_l$ is strictly increasing.

For one contrast, either nonzero multiplier forces every higher harmonic of
the group aggregate to vanish.  For two contrasts, the determinant of the
coefficient matrix is
\begin{equation}
 b_l(\alpha_1)b_l(\alpha_2)
 [R_l(\alpha_2)-R_l(\alpha_1)]\neq0,
 \label{eq:two-contrast-determinant}
\end{equation}
so both group harmonic moments vanish separately.  Mass and barycenter then
complete Eqs.~\eqref{eq:one-contrast-equality} and
\eqref{eq:two-contrast-equality}.

If $K\geq3$, let $\mathsf M_l$ be the $2\times K$ matrix whose $a$th column
is $(a_l(\alpha_a),b_l(\alpha_a))^T$.  It has rank two and a nontrivial
nullspace.  Choose a nonzero real null vector
$(c_1,\ldots,c_K)$ and a real spherical harmonic $Y_{lm}$ with $l\geq2$.
For sufficiently small nonzero $\varepsilon$, assign to every node in group
$G_a$ the probability density
\begin{equation}
 1+\varepsilon\frac{c_a}{|G_a|}Y_{lm}(\bu)
 \label{eq:hetero-construction-density}
\end{equation}
with respect to $\omega$.  It is nonnegative, normalized, and has zero first
moment.  Only the selected degree is perturbed, and the two nullspace
equations make both total tensor conditions vanish.  This constructs the
claimed anisotropic minimizers and completes the proof of
Theorem~\ref{thm:heterogeneous-fixed}.

\section{Pure-state endpoint}
\label{app:pure}

\subsection{Minimax value and tensor rigidity}

For $\br=s\bu$ and $c=\br\cdot\bn=st$, the pure tangent-trace integrand
satisfies, whenever $1+c>0$,
\begin{equation}
 \frac{s^2-c^2}{1+c}
 =
 (s-c)\frac{s+c}{1+c}
 \leq s-c,
 \label{eq:pure-pointwise-integrand}
\end{equation}
because $s\leq1$.  At the omitted point $s=1,c=-1$, the Fisher contribution
is zero and the same upper bound remains valid.  Since
$\int\br\,\dd\mu=0$,
\begin{equation}
 \tau_{\mu}^{\mathrm{pure}}(\bn)
 \leq
 N\int_{\Bbar}\|\br\|\,\mu(\dd\br)
 \leq N
 \quad\text{for every }\bn.
 \label{eq:pure-pointwise-trace-upper}
\end{equation}
The rotational average is the endpoint of
Eq.~\eqref{eq:average-trace-identity}:
\begin{equation}
 \int_{S^2}\tau_{\mu}^{\mathrm{pure}}(\bn)\omega(\dd\bn)
 =
 2N\int_{\Bbar}f(\|\br\|)\mu(\dd\br)
 \leq N.
 \label{eq:pure-average}
\end{equation}
The trace-balanced argument gives the lower bound
Eq.~\eqref{eq:pure-minimax}; the pure Haar POVM has Fisher field
$(N/2)P_{\bn}$ and attains it.

Suppose equality holds.  The loss bound implies
$\tau_{\mu}^{\mathrm{pure}}(\bn)\geq N$ at every direction, while
Eq.~\eqref{eq:pure-pointwise-trace-upper} gives the reverse inequality.
Thus the trace is identically $N$.  Equality in the rotational average forces
$\|\br\|=1$ almost everywhere.  For a zero-barycenter unit-sphere measure,
\begin{align}
 \frac1N\tau_{\mu}^{\mathrm{pure}}(\bn)
 &=
 \int_{\bu\neq-\bn}(1-\bu\cdot\bn)\mu(\dd\bu)
 \nonumber\\
 &=
 1-2\mu(\{-\bn\}).
 \label{eq:pure-trace-atoms}
\end{align}
Hence equality forces $\mu$ to be atomless.  Equality in trace balance also
forces
\begin{equation}
 F_{\mu}^{\mathrm{pure}}(\bn)=\frac N2P_{\bn}
 \quad\text{for every }\bn.
 \label{eq:pure-isotropic-field}
\end{equation}

The scalar trace cannot identify $\mu$ at this endpoint.  Define
\begin{equation}
 h_1(t)=(1+t)\log(1+t),
 \qquad h_1(-1)=0,
 \label{eq:pure-potential}
\end{equation}
and, for $\delta=\mu-\omega$,
\begin{equation}
 H_{\delta}(\bn)
 =
 \int_{S^2}h_1(\bu\cdot\bn)\delta(\dd\bu).
 \label{eq:pure-H-delta}
\end{equation}
Although $h_1$ is not $C^2$ at $t=-1$, the identity
\eqref{eq:TF-Hessian-identity} remains valid in the distributional sense at
$\alpha=1$.  To verify this, fix $\bu\in S^2$ and let
$r=d_{S^2}(\bn,-\bu)$.  As $r\downarrow0$,
\begin{align}
 1+\bu\cdot\bn
 &=
 1-\cos r
 =
 \frac{r^2}{2}+O(r^4),
 \label{eq:pure-antipodal-distance}\\
 h_1(\bu\cdot\bn)
 &=
 O\!\left(r^2|\log r|\right),
 \label{eq:pure-potential-local-bound}\\
 \left\|\nabla_{\bn}h_1(\bu\cdot\bn)\right\|
 &=
 O\!\left(r|\log r|\right).
 \label{eq:pure-gradient-local-bound}
\end{align}
Apply integration by parts on
$S^2\setminus B_{\varepsilon}(-\bu)$.  The boundary terms involving the
first derivative and the function itself are respectively
$O(\varepsilon^2|\log\varepsilon|)$ and
$O(\varepsilon^3|\log\varepsilon|)$, and therefore vanish as
$\varepsilon\downarrow0$.  Thus the distributional Hessian contains no
point-supported term at $-\bu$.  Away from that point, the ordinary zonal
Hessian formula gives
\begin{equation}
 \TF\nabla_{\bn}^2h_1(\bu\cdot\bn)
 =
 \TF\left[
  \frac{P_{\bn}\bu\bu^TP_{\bn}}
  {1+\bu\cdot\bn}
 \right].
 \label{eq:pure-distributional-Hessian}
\end{equation}
The tensor on the right is uniformly bounded, since
\begin{equation}
 \left\|
  \frac{P_{\bn}\bu\bu^TP_{\bn}}
  {1+\bu\cdot\bn}
 \right\|
 =
 \frac{1-(\bu\cdot\bn)^2}{1+\bu\cdot\bn}
 =
 1-\bu\cdot\bn
 \leq2.
 \label{eq:pure-tensor-bound}
\end{equation}
Moreover, the preceding trace argument has already shown that $\mu$, and
hence $\delta=\mu-\omega$, has no atoms.  The omission of the outcome
$\bu=-\bn$ in the ordinary-Fisher convention therefore does not change the
integral.  Fubini's theorem now yields
\begin{equation}
 \TF\nabla^2H_{\delta}(\bn)
 =
 \TF\mathcal F_{1,\delta}(\bn).
 \label{eq:pure-integrated-Hessian}
\end{equation}
Since both $\mu$ and the Haar measure have the isotropic Fisher field in
Eq.~\eqref{eq:pure-isotropic-field}, the right-hand side vanishes, and hence
\begin{equation}
 \TF\nabla^2H_{\delta}=0.
 \label{eq:pure-TF-zero}
\end{equation}

Taking $\alpha\uparrow1$ in Eq.~\eqref{eq:b-closed} by dominated
convergence gives, for every $l\geq2$,
\begin{equation}
 \int_{-1}^{1}h_1(t)P_l(t)\,\dd t
 =
 \frac{(-1)^l}{2^ll(l-1)}
 \int_{-1}^{1}
 \frac{(1-t^2)^l}{(1+t)^{l-1}}\,\dd t
 \neq0.
 \label{eq:pure-potential-spectrum}
\end{equation}
The trace-free Hessian is injective on every scalar harmonic degree
$l\geq2$, so Eq.~\eqref{eq:pure-TF-zero} annihilates all higher harmonic
moments of $\delta$.  Mass and barycenter annihilate degrees zero and one.
Therefore $\delta=0$, proving pure Haar rigidity.

\subsection{Finite-support lower bound and convergence}

The radial promotion in Eqs.~\eqref{eq:radial-promotion} and
\eqref{eq:radial-Loewner} remains valid at $\eta=1$, with both sides assigned
zero contribution at an exactly antipodal atom.  For a promoted unit design
$\nu=\sum_ap_a\delta_{\bu_a}$ and an atom of maximal weight $p$,
Eq.~\eqref{eq:pure-trace-atoms} at $-\bu_k$ gives
\begin{equation}
 \tau_{\nu}^{\mathrm{pure}}(-\bu_k)=N(1-2p)
 \leq N\frac{M-2}{M}.
 \label{eq:pure-atom-bound}
\end{equation}
This proves Eq.~\eqref{eq:pure-M-general-lower}; the support argument for
$M\leq3$ is unchanged.

To prove convergence, partition $S^2$ into $L$ equal-area regions of maximal
diameter $\delta_L\to0$ and choose one representative $\bu_a$ in each region
\cite{Leopardi2006}.  Symmetrize the empirical measure,
\begin{equation}
 \mu_{2L}
 =
 \frac1{2L}\sum_{a=1}^{L}
 (\delta_{\bu_a}+\delta_{-\bu_a}).
 \label{eq:pure-symmetric-quadrature}
\end{equation}
It has zero barycenter, at most $2L$ atoms, and converges weakly to Haar.
Since the finite-support classes are nested, convergence along this even
subsequence implies the full limit in $M$.  For fixed $\bn$, write
\begin{equation}
 K_{\bn}(\bu)
 =
 \frac{P_{\bn}\bu\bu^TP_{\bn}}{1+\bu\cdot\bn},
 \qquad \bu\neq-\bn,
 \label{eq:pure-kernel}
\end{equation}
and assign the ordinary-Fisher value zero at $\bu=-\bn$.  The operator norm
of $K_{\bn}$ is $1-\bu\cdot\bn\leq2$.  On the complement of an angular
$r$-cap about $-\bn$, the family is uniformly Lipschitz with constant
$O(r^{-1})$.  Cells intersecting that cap have total weight
$O((r+\delta_L)^2)$.  Consequently the quadrature error is bounded uniformly
in $\bn$ by
\begin{equation}
 O((r+\delta_L)^2)+O(\delta_L/r).
 \label{eq:pure-quadrature-error}
\end{equation}
Taking $r=\sqrt{\delta_L}$ shows that the Fisher fields converge uniformly to
$(N/2)P_{\bn}$.  Continuity of every trace-balanced loss on a compact positive
spectral interval proves Eq.~\eqref{eq:pure-M-limit-general}.

\subsection{The pure tetrahedron and its uniqueness}

Away from the four antipodes, the mixed tetrahedral formulas have a regular
limit at $\eta=1$.  Equation~\eqref{eq:tetra-A-gap} becomes nonnegative
because, for $q<0$ and $a=3\sqrt3|q|$,
\begin{equation}
 1+3s+6\sqrt3q
 \geq
 1+a^{4/3}-2a\geq0,
 \label{eq:pure-tetra-A-gap}
\end{equation}
with equality only at $a=1$.  The determinant inequality
\eqref{eq:D-det-target} has the same endpoint by
Eq.~\eqref{eq:D-gap-lower}.  At an antipode, the antipodal outcome is omitted
and the three remaining projected tetrahedral vectors give
\begin{equation}
 F_{\mathrm{tet}}^{\mathrm{pure}}(-\bu_k)=\frac N4P_{-\bu_k}.
 \label{eq:pure-tetra-antipode-F}
\end{equation}
This proves the pure tetrahedral values and worst-direction set.

It remains to prove uniqueness, because strict concavity is lost at
$\eta=1$.  Equality in the four-atom lower bound first forces four equal unit
weights and four unit vectors satisfying
\begin{equation}
 \sum_{a=1}^{4}\bu_a=0,
 \qquad
 \Span\{\bu_a\}=\R^3.
 \label{eq:pure-four-frame}
\end{equation}
Let $U$ be the $4\times3$ matrix with rows $\bu_a^T$, put
\begin{equation}
 S=U^TU,
 \qquad
 \bv_a=4S^{-1}\bu_a.
 \label{eq:pure-dual-vectors}
\end{equation}
Because $\ker U^T=\Span\{\bone\}$, the matrix
$U(4S^{-1})U^T$ is four times the orthogonal projector onto
$\bone^{\perp}$, so
\begin{equation}
 \bu_k\cdot\bv_i=4\delta_{ki}-1.
 \label{eq:dual-Gram}
\end{equation}

At $\bn=-\bu_k$, the remaining three outcomes form a saturated regular
classical model on the tangent plane.  The vectors
$P_{-\bu_k}\bv_i$, $i\neq k$, are the unique locally unbiased estimator
values: globally,
\begin{equation}
 \sum_{i=1}^{4}\frac{1+\bu_i\cdot\bn}{4}\,\bv_i=\bn,
 \label{eq:ambient-unbiased}
\end{equation}
and tangent differentiation gives the identity map.  Therefore their tangent
covariance equals the inverse Fisher matrix.  For one copy,
\begin{equation}
 A_k
 :=
 \Tr[F^{(1)}(-\bu_k)^{-1}]
 =
 \frac14\sum_{i\neq k}
 (1-\bu_i\cdot\bu_k)(\|\bv_i\|^2-1).
 \label{eq:pure-Ak}
\end{equation}
Summing over $k$ and using
$\sum_{k\neq i}(1-\bu_i\cdot\bu_k)=4$ gives
\begin{equation}
 \frac14\sum_{k=1}^{4}A_k
 =
 4\Tr(S^{-1})-1.
 \label{eq:pure-average-Ak}
\end{equation}
Since $\Tr S=4$, the arithmetic--harmonic mean inequality gives
\begin{equation}
 \Tr(S^{-1})\geq\frac94,
 \label{eq:S-inverse-bound}
\end{equation}
with equality if and only if $S=(4/3)I_3$.  Hence
$\max_kA_k\geq8$ for one copy, and equality forces
\begin{equation}
 UU^T
 =
 \frac43\left(I_4-\frac14\bone\bone^T\right),
 \label{eq:pure-Gram-tetra}
\end{equation}
whose off-diagonal entries are $-1/3$.  This proves $A$-loss uniqueness.
For the $D$ loss, equality at each antipode in the trace lower bound forces
the two tangent eigenvalues to be equal; hence $A_k=8$ at every antipode and
the same argument applies.  Theorem~\ref{thm:pure} follows.

\section{Separable, LOCC, and collective bounds}
\label{app:architectures}

\subsection{A finite-copy trace bound for separable POVMs}

First consider a finite-outcome separable POVM.  Every positive separable
effect in finite dimension is a finite sum of rank-one product effects;
adjoining the summand label gives a POVM refinement, and recording a
refinement can only increase classical Fisher information.  For a standard
Borel outcome space, apply the finite-outcome argument to every finite
measurable partition.  The Fisher matrix of the corresponding coarse
graining is the second moment of the conditional expectation of the score.
Along an increasing sequence of generating partitions these matrices
converge monotonically to the Fisher matrix of the full experiment by the
$L^2$ martingale convergence theorem.  It is therefore sufficient to prove
the bound for a finite refined POVM, which can be written as
\begin{equation}
 M(\dd\omega)
 =
 \bigotimes_{j=1}^{N}
 (\I+\bu_j(\omega)\cdot\bsigma)\,
 \nu(\dd\omega),
 \qquad \|\bu_j(\omega)\|=1.
 \label{eq:product-refinement}
\end{equation}
Here $\nu=2^{-N}\Tr M$ is a probability measure.  Expanding the normalization
$\int M=\I^{\otimes N}$ in Pauli tensors gives, for every nonempty set of
distinct nodes $j_1,\ldots,j_r$,
\begin{equation}
 \int
 \bu_{j_1}\otimes\cdots\otimes\bu_{j_r}
 \,\nu(\dd\omega)=0.
 \label{eq:mixed-product-moments}
\end{equation}

Put $c_j=\bu_j\cdot\bn$.  The outcome density and tangent score are
\begin{align}
 p(\omega|\bn)
 &=
 \prod_{j=1}^{N}(1+\eta c_j),
 \label{eq:product-probability}\\
 s_{\bv}(\omega|\bn)
 &=
 \eta\sum_{j=1}^{N}
 \frac{\bu_j\cdot\bv}{1+\eta c_j}.
 \label{eq:product-score}
\end{align}
When the tangent trace of $\mathbb E(ss^T)$ is expanded, every cross-node
term vanishes by Eq.~\eqref{eq:mixed-product-moments}.  After expanding
$\prod_{k\neq j,l}(1+\eta c_k)$, both
$\bu_j\cdot\bu_l$ and $c_jc_l$ contract a nonempty mixed product moment.
Only diagonal terms remain:
\begin{equation}
 \Tr F_{\mathrm{sep}}(\bn)
 =
 \eta^2\sum_{j=1}^{N}
 \int
 \frac{1-c_j^2}{1+\eta c_j}
 \prod_{k\neq j}(1+\eta c_k)\,\dd\nu.
 \label{eq:separable-trace-diagonal}
\end{equation}
Using
\begin{equation}
 \frac{1-c^2}{1+\eta c}
 =
 1-\eta c
 -\frac{(1-\eta^2)c^2}{1+\eta c}
 \label{eq:separable-scalar-identity}
\end{equation}
and Eq.~\eqref{eq:mixed-product-moments} once more gives the exact identity
\begin{multline}
 \Tr F_{\mathrm{sep}}(\bn)
 =N\eta^2
 \\
 {}
 -\eta^2(1-\eta^2)
 \sum_{j=1}^{N}
 \int
 \frac{c_j^2}{1+\eta c_j}
 \prod_{k\neq j}(1+\eta c_k)\,\dd\nu.
 \label{eq:separable-trace-identity}
\end{multline}
At a fixed $\bn$, equality with $N\eta^2$ holds if and only if
$c_j=0$ for every node and $\nu$-almost every refined outcome.

For all $c_j\in[-1,1]$,
\begin{equation}
 \frac1{1+\eta c_j}
 \prod_{k\neq j}(1+\eta c_k)
 \geq
 \frac{(1-\eta)^{N-1}}{1+\eta}.
 \label{eq:separable-factor-lower}
\end{equation}
Averaging Eq.~\eqref{eq:separable-trace-identity} over $\bn$ and using
$\int_{S^2}(\bu_j\cdot\bn)^2\omega(\dd\bn)=1/3$ yields
\begin{equation}
 \int_{S^2}\Tr F_{\mathrm{sep}}(\bn)\omega(\dd\bn)
 \leq
 N\eta^2\left[1-\frac{(1-\eta)^N}{3}\right].
 \label{eq:separable-average-bound}
\end{equation}
At a direction where the trace does not exceed this average,
$\Tr(F^{-1})\geq4/\Tr F$.  Coarse graining cannot increase Fisher
information, so the bound applies to the original separable POVM and hence to
all LOCC subclasses.  This proves
Eq.~\eqref{eq:hierarchy-separable-lower}.

\subsection{One-way LOCC attainability}

Let $m=\lceil N^{2/3}\rceil$.  Measure approximately $m/3$ pilot copies along
each Cartesian Pauli axis.  If the three sample means are divided by $\eta$,
collected into $\widetilde\bn$, and radially projected to a unit vector
$\widehat\bn$, then uniformly in $\bn$,
\begin{equation}
 \mathbb E_{\bn}\|\widehat\bn-\bn\|
 \leq
 2\mathbb E_{\bn}\|\widetilde\bn-\bn\|
 \leq
 \frac{C_{\eta}}{\sqrt m}.
 \label{eq:pilot-error}
\end{equation}
The first inequality follows from nearest radial projection to $S^2$, and
the second from the bounded variances of the three sample means.

Conditional on $\widehat\bn$, measure each remaining copy along a Haar-random
projective line in $\widehat\bn^{\perp}$ and retain the line and binary
outcome.  Its conditional one-copy Fisher operator at the true $\bn$ is
\begin{equation}
 K_{\eta}(\bn,\widehat\bn)
 =
 \eta^2
 \int_{\mathbb P(\widehat\bn^{\perp})}
 \frac{P_{\bn}\bu\bu^TP_{\bn}}
 {1-\eta^2(\bu\cdot\bn)^2}
 \,h_{\widehat\bn}(\dd\ell),
 \label{eq:adaptive-kernel}
\end{equation}
where either orientation $\bu$ of $\ell$ gives the same integrand.  Since the
denominator is bounded below by $1-\eta^2$, coupling the two projective
circles by a rotation carrying $\bn$ to $\widehat\bn$ gives the uniform
Lipschitz estimate
\begin{equation}
 \left\|
  K_{\eta}(\bn,\widehat\bn)-\frac{\eta^2}{2}P_{\bn}
 \right\|
 \leq
 C'_{\eta}\|\widehat\bn-\bn\|.
 \label{eq:adaptive-Lipschitz}
\end{equation}
The Fisher chain rule for a sequential recorded experiment states that the
total Fisher information equals the pilot information plus the expected
conditional Fisher information of the second stage.  Equations
\eqref{eq:pilot-error} and \eqref{eq:adaptive-Lipschitz} therefore imply,
uniformly in $\bn$,
\begin{equation}
 F_N(\bn)
 \succeq
 \left[
  \frac{(N-m)\eta^2}{2}
  -C''_{\eta}\frac{N-m}{\sqrt m}
 \right]P_{\bn}.
 \label{eq:adaptive-Fisher-lower}
\end{equation}
With $m\asymp N^{2/3}$,
\begin{equation}
 \sup_{\bn}\Tr[F_N(\bn)^{-1}]
 \leq
 \frac4{N\eta^2}[1+O(N^{-1/3})].
 \label{eq:adaptive-A-upper}
\end{equation}
Together with Eq.~\eqref{eq:hierarchy-separable-lower}, this proves
Eqs.~\eqref{eq:LOCC-asymptotic} and \eqref{eq:SEP-asymptotic}.  The
construction is one-way LOCC because only the pilot record controls later
single-copy settings.  It is the Fisher-information analogue of standard
two-stage adaptive state-estimation schemes
\cite{GillMassar2000,Bagan2005,HayashiMatsumoto2008}.

\subsection{Collective lower bound}

At $\bn=(0,0,1)$, choose tangent coordinates $x,y$. The symmetric logarithmic
derivatives (SLDs) are
\begin{equation}
 L_x=\eta\sigma_x,
 \qquad
 L_y=\eta\sigma_y,
 \label{eq:SLDs}
\end{equation}
so the SLD Fisher matrix is $J=\eta^2I_2$.  The locally unbiased operators
$X_x=\sigma_x/\eta$ and $X_y=\sigma_y/\eta$ have complex covariance matrix
\begin{equation}
 Z=
 \begin{pmatrix}
  \eta^{-2}&\ii\eta^{-1}\\
  -\ii\eta^{-1}&\eta^{-2}
 \end{pmatrix}.
 \label{eq:Holevo-Z}
\end{equation}
The fixed-radius qubit model is $D$ invariant, so the Holevo bound for unit
weight is
\begin{equation}
 C_H
 =
 \Tr\Re Z+\Tr|\Im Z|
 =
 \frac{2(1+\eta)}{\eta^2}
 \label{eq:Holevo-value}
\end{equation}
\cite{Holevo2011,Suzuki2016,Demkowicz2020}. For $N$ copies, the SLDs are the sums of their one-copy counterparts, so
$J_N=NJ$, and the corresponding commutator matrix also scales linearly with
$N$.  The $D$-invariant Holevo formula therefore gives
$C_H^{(N)}=C_H/N$.

To apply the bound to a classical Fisher experiment, fix a regular point and
let $s(y)$ be the score of an arbitrary collective POVM.  The random vector
$F^{-1}s(y)$ is locally unbiased and has covariance $F^{-1}$.  The Holevo
inequality therefore gives Eq.~\eqref{eq:hierarchy-collective-lower} directly
for every collective measurement.

\subsection{Two-copy collective upper bounds}

A tetrahedral qubit SIC is a complex projective $2$-design.  Writing
$\Pi_+$ for the projector onto the symmetric two-qubit subspace, \begin{equation}
 \frac34\sum_{a=1}^{4}
 (|\bu_a\rangle\langle\bu_a|)^{\otimes2}
 =\Pi_+.
 \label{eq:collective-normalization}
\end{equation}
Together with $M_-=\Pi_-$, this proves POVM normalization.  The five outcome
probabilities are
\begin{align}
 p_a(\bn)
 &=
 \frac3{16}(1+\eta\bu_a\cdot\bn)^2,
 \label{eq:collective-probabilities}\\
 p_-&=\frac{1-\eta^2}{4}.
 \label{eq:singlet-probability}
\end{align}
The singlet carries no directional information.  For
$\bv\in T_{\bn}S^2$,
\begin{align}
 \langle\bv,F_{\mathrm{collSIC}}^{(2)}(\bn)\bv\rangle
 &=
 \frac{3\eta^2}{4}
 \sum_{a=1}^{4}(\bu_a\cdot\bv)^2
 \nonumber\\
 &=
 \eta^2\|\bv\|^2,
 \label{eq:collective-Fisher-proof}
\end{align}
because $\sum_a\bu_a\bu_a^T=(4/3)I_3$.  This proves
Eq.~\eqref{eq:collective-SIC-Fisher}.  Independent use on $m$ pairs gives
$m\eta^2P_{\bn}$; an additional single-copy Haar POVM contributes
$f(\eta)P_{\bn}$.  Equations~\eqref{eq:collective-even-upper} and
\eqref{eq:collective-odd-upper} follow.

\subsection{An asymptotically sharp Schur-coherent POVM}

We now construct a collective POVM whose ordinary Fisher information attains
the Holevo coefficient.  Write
\begin{equation}
 \eta=\tanh\beta,
 \qquad
 \bm J=\frac12\sum_{k=1}^{N}\bsigma^{(k)}.
 \label{eq:beta-total-spin}
\end{equation}
The Schur--Weyl decomposition is
\begin{equation}
 (\mathbb C^2)^{\otimes N}
 =
 \bigoplus_{j\in\mathcal J_N}
 \left(\mathcal H_j\otimes\mathbb C^{m_{N,j}}\right),
 \label{eq:Schur-decomposition}
\end{equation}
where $\mathcal J_N=\{0,1,\ldots,N/2\}$ for even $N$ and
$\mathcal J_N=\{1/2,3/2,\ldots,N/2\}$ for odd $N$, and $\mathcal H_j$ is
the spin-$j$ irreducible representation.  Its multiplicity is
\begin{equation}
 m_{N,j}
 =
 \binom{N}{N/2-j}-\binom{N}{N/2-j-1},
 \label{eq:Schur-multiplicity}
\end{equation}
with out-of-range binomial coefficients set to zero.  Let
$J_{\bn}^{(j)}$ denote the spin-$j$ angular-momentum component along $\bn$.
In this decomposition
\begin{equation}
 \rho_{\bn}^{\otimes N}
 =
 \frac1{(2\cosh\beta)^N}
 \bigoplus_{j\in\mathcal J_N}
 \left(
  \ee^{2\beta J_{\bn}^{(j)}}\otimes I_{m_{N,j}}
 \right).
 \label{eq:Schur-state}
\end{equation}
Such sector decompositions and covariant measurements are standard in
collective qubit estimation \cite{KeylWerner2001,BaganFull2006}.

In every spin-$j$ sector, let $|j,\bu\rangle$ be the spin-coherent state in
direction $\bu$ and perform the covariant POVM
\begin{equation}
 M_j(\dd\bu)
 =
 (2j+1)|j,\bu\rangle\langle j,\bu|
 \,\omega(\dd\bu)\otimes I_{m_{N,j}}.
 \label{eq:Schur-coherent-POVM}
\end{equation}
The effects are embedded in their corresponding Schur blocks.  Since
\begin{equation}
 (2j+1)\int_{S^2}
 |j,\bu\rangle\langle j,\bu|\,\omega(\dd\bu)
 =I_{\mathcal H_j},
 \label{eq:coherent-resolution}
\end{equation}
the collection of all sector outcomes $(j,\bu)$ is a
parameter-independent POVM.

The sector probability is independent of $\bn$ and equals
\begin{equation}
 q_{N,j}
 =
 \frac{m_{N,j}}{(2\cosh\beta)^N}
 \frac{\sinh[(2j+1)\beta]}{\sinh\beta}.
 \label{eq:Schur-sector-probability}
\end{equation}
Conditional on the sector, put $r=2j$.  The coherent-state identity
\begin{equation}
 \langle j,\bu|\ee^{2\beta J_{\bn}^{(j)}}|j,\bu\rangle
 =
 (\cosh\beta)^r(1+\eta\bu\cdot\bn)^r
 \label{eq:coherent-exponential}
\end{equation}
gives the normalized directional density
\begin{equation}
 p_j(\bu|\bn)
 =
 \frac{2(1+\eta\bu\cdot\bn)^r}{I_r(\eta)},
 \label{eq:Schur-conditional-density}
\end{equation}
where
\begin{align}
 I_r(\eta)
 &=
 \int_{-1}^{1}(1+\eta t)^r\,\dd t
 \nonumber\\
 &=
 \frac{(1+\eta)^{r+1}-(1-\eta)^{r+1}}
 {\eta(r+1)}.
 \label{eq:Ir-def}
\end{align}
For a tangent vector $\bv$,
\begin{equation}
 \dd\log p_j(\bu|\bn)[\bv]
 =
 \frac{r\eta\,\bu\cdot\bv}{1+\eta\bu\cdot\bn}.
 \label{eq:Schur-score}
\end{equation}
Rotational covariance therefore gives
\begin{equation}
 F_j(\bn)=\lambda_r(\eta)P_{\bn},
 \label{eq:Schur-sector-Fisher}
\end{equation}
with
\begin{equation}
 \lambda_r(\eta)
 =
 \frac{r^2\eta^2}{2I_r(\eta)}
 \int_{-1}^{1}
 (1-t^2)(1+\eta t)^{r-2}\,\dd t.
 \label{eq:lambda-r-integral}
\end{equation}
Here $\lambda_0=0$ and $\lambda_1=f(\eta)$.  For $r\geq2$, integration by
parts gives the exact formula
\begin{equation}
 \lambda_r(\eta)
 =
 \frac{r^2}{r-1}
 \left(1-\frac{I_{r-1}(\eta)}{I_r(\eta)}\right).
 \label{eq:lambda-r-exact}
\end{equation}
Indeed, the vanishing boundary term in the derivative of
$(1-t^2)(1+\eta t)^{r-1}$ yields
\begin{equation}
 \int_{-1}^{1}(1-t^2)(1+\eta t)^{r-2}\,\dd t
 =
 \frac{2[I_r(\eta)-I_{r-1}(\eta)]}{(r-1)\eta^2}.
 \label{eq:lambda-integration-parts}
\end{equation}

Because the sector label carries no directional information, the total
Fisher field is exactly
\begin{equation}
 F_{\mathrm{Schur}}^{(N)}(\bn)
 =
 \kappa_N(\eta)P_{\bn},
 \qquad
 \kappa_N(\eta)
 =
 \sum_jq_{N,j}\lambda_{2j}(\eta).
 \label{eq:Schur-total-Fisher}
\end{equation}
Its $A$ loss is $2/\kappa_N(\eta)$.

It remains to evaluate $\kappa_N$.  Set
\begin{equation}
 x=\frac{1-\eta}{1+\eta}\in(0,1).
 \label{eq:x-eta}
\end{equation}
For $r\geq2$,
\begin{equation}
 \frac{I_{r-1}(\eta)}{I_r(\eta)}
 =
 \frac{r+1}{r(1+\eta)}
 \frac{1-x^r}{1-x^{r+1}},
 \label{eq:Ir-ratio}
\end{equation}
and hence
\begin{multline}
 \lambda_r(\eta)-\frac{\eta r}{1+\eta}
 =
 -\frac{r(1-\eta)}{(r-1)(1+\eta)}
 \\
 {}
 +\frac{r(r+1)}{(r-1)(1+\eta)}
 \frac{(1-x)x^r}{1-x^{r+1}}.
 \label{eq:lambda-linear-difference}
\end{multline}
The first term is uniformly bounded in $r$, while the second is bounded
because $rx^r$ is bounded for fixed $x<1$.  Absorbing the cases $r=0,1$,
therefore,
\begin{equation}
 \lambda_r(\eta)
 =
 \frac{\eta r}{1+\eta}+O_{\eta}(1)
 \quad\text{uniformly for }r\geq0.
 \label{eq:lambda-linear-asymptotic}
\end{equation}

Regard $j$ as a random variable with probabilities $q_{N,j}$ and define the
sector-label operator
\begin{equation}
 \widehat J
 =
 \bigoplus_j
 jI_{\mathcal H_j}\otimes I_{m_{N,j}}.
 \label{eq:Jhat}
\end{equation}
In every spin sector, $jI\geq J_{\bn}^{(j)}$, so
\begin{equation}
 \mathbb E[j]
 \geq
 \Tr(\rho_{\bn}^{\otimes N}J_{\bn})
 =
 \frac{N\eta}{2}.
 \label{eq:mean-j-lower}
\end{equation}
Moreover, $\widehat J(\widehat J+I)=\bm J^2$, and product-state moments give
\begin{equation}
 \mathbb E[j(j+1)]
 =
 \Tr(\rho_{\bn}^{\otimes N}\bm J^2)
 =
 \frac{3N}{4}+\frac{N(N-1)\eta^2}{4}.
 \label{eq:mean-j-square}
\end{equation}
Writing $s_N=\mathbb E[j]$, Jensen's inequality yields
\begin{align}
 s_N
 &\leq
 \frac{-1+\sqrt{N^2\eta^2+N(3-\eta^2)+1}}{2}
 \nonumber\\
 &=
 \frac{N\eta}{2}+O_{\eta}(1).
 \label{eq:mean-j-upper}
\end{align}
Together with Eq.~\eqref{eq:mean-j-lower}, this proves
\begin{equation}
 \mathbb E[j]=\frac{N\eta}{2}+O_{\eta}(1).
 \label{eq:mean-j-asymptotic}
\end{equation}

Using Eq.~\eqref{eq:lambda-linear-asymptotic} in
Eq.~\eqref{eq:Schur-total-Fisher},
\begin{align}
 \kappa_N(\eta)
 &=
 \frac{2\eta}{1+\eta}\mathbb E[j]+O_{\eta}(1)
 \nonumber\\
 &=
 \frac{N\eta^2}{1+\eta}+O_{\eta}(1).
 \label{eq:Schur-kappa-asymptotic}
\end{align}
Consequently,
\begin{equation}
 \frac{2}{\kappa_N(\eta)}
 =
 \frac{2(1+\eta)}{N\eta^2}+O_{\eta}(N^{-2}).
 \label{eq:Schur-A-asymptotic}
\end{equation}
For sufficiently large $N$, the remainder is bounded above by
$K_{\eta}/N^2$, proving Eq.~\eqref{eq:collective-sharp-upper}.  Combining
this explicit upper bound with Eq.~\eqref{eq:hierarchy-collective-lower}
proves Eq.~\eqref{eq:COLL-asymptotic}.

At $\eta=1$, the pure-state Holevo bound is $4/N$ \cite{Hayashi1998}, and the
fixed local Haar POVM already attains it.  The inclusions in
Eq.~\eqref{eq:architecture-inclusion} prove
Eq.~\eqref{eq:pure-hierarchy-collapse}.

\bibliography{ref}

\end{document}